\documentclass[pra,twocolumn,noshowkeys,superscriptaddress,preprintnumbers,amsmath,amssymb, showpacs,nofootinbib]{revtex4}
\usepackage{verbatim}

\def\identity{\leavevmode\hbox{\small1\kern-3.8pt\normalsize1}}

\newcommand{\trace}[2][]{\textsf{tr}_{#1}\left(#2\right)}

\newcommand{\abs}[1]{\left\vert #1 \right\vert}

\newcommand{\half}{\frac{1}{2}}

\newcommand{\bra}[1]{\langle #1 \vert}
\newcommand{\ket}[1]{\vert #1 \rangle}

\newcommand{\branoopket}[2]{\langle #1 \vert #2 \rangle}
\newcommand{\ketbra}[2]{\vert {#1} \rangle\langle {#2} \vert}

\newcommand{\norm}[1]{\left\| #1 \right\|}

\usepackage{amsmath}
\usepackage{bbm,amsfonts}

\newcommand{\hide}[1]{}
\newcommand{\vect}[1]{\mathbf{#1}}

\renewcommand{\epsilon}{\varepsilon}

\def\Id{{\mathbbm 1}}

\def\C{{\mathcal C}}
\def\R{{\mathcal R}}
\def\M{{\mathcal M}}
\def\H{{\mathcal H}}
\def\E{{\mathcal E}}

\def\U{{\mathcal U}}

\usepackage{amsthm}
\theoremstyle{definition}

\usepackage{graphicx}

\usepackage{amsmath,enumerate}

\usepackage{amsmath}
\usepackage{bbm,amsfonts}
\def\Id{{\mathbbm 1}}

\begin{document}

\newcommand{\affilmpq}{\affiliation{Max-Planck-Institut f{\"{u}}r
Quantenoptik,
Hans-Kopfermann-Str.\ 1, D-85748 Garching, Germany.}}

\title{Limitations of Passive Protection of Quantum Information}
\author{Fernando Pastawski}
\affilmpq
\author{Alastair Kay}
\affiliation{Centre for Quantum Computation,
             DAMTP,
             Centre for Mathematical Sciences,
             University of Cambridge,
             Wilberforce Road,
             Cambridge CB3 0WA, UK}
\affilmpq
\author{Norbert Schuch}
\affilmpq
\author{Ignacio Cirac}
\affilmpq

\begin{abstract}
The ability to protect quantum information from the effect of noise is one of the major goals of quantum information processing.
In this article, we study limitations on the asymptotic stability of quantum information stored in passive $N$-qubit systems.
We consider the effect of small imperfections in the implementation of the protecting Hamiltonian in the form of perturbations or weak coupling to a ground state environment.
We prove that, regardless of the protecting Hamiltonian, there exists a perturbed evolution that necessitates a final error correcting step when the state of the memory is read.
Such an error correction step is shown to require a finite error threshold, the lack thereof being exemplified by the 3D compass model.
We go on to present explicit weak Hamiltonian perturbations which destroy the logical information stored in the 2D toric code in a time $O(\log(N))$.
 \end{abstract}

\pacs{03.67.Lx, 03.67.Pp}

\date{\today}

\maketitle

\setcounter{tocdepth}{3}
\tableofcontents

\section{\label{sec:Introduction}
Introduction}
Quantum information processing promises exciting new capabilities for a host of computational \cite{shor_1997,childs_quantum_2008,harrow_quantum_2009} and cryptographic \cite{bennet_brassard_1984,ekert_91} tasks, if only we can fabricate devices that take advantage of the subtle and very fragile effects of quantum mechanics.
The theory of quantum error-correcting codes (QECCs) and fault-tolerance \cite{shor_scheme_1995, aharonov_fault_1996, gottesman_theory_1998, gottesman_introduction_2009} assure that this fragility can be overcome at a logical level once an error rate per element below a certain threshold is achieved.
However, providing a scalable physical implementation of computational elements with the required degree of precision and control has proven to be a task of extreme difficulty.
Thus, one might hope to design superior fault-tolerant components whose robustness is enforced in a more natural way at a physical level.

A first step in this daunting task is to concentrate not on universal quantum computation, but on one sub-protocol within this; the storage of quantum information.
Thus, the aim is to find systems naturally assuring the stability of quantum information, just like magnetic domains in a hard disk provide stable storage of classical information.
The quest for such a passive quantum memory was pioneered by Kitaev \cite{kitaev_fault-tolerant_2003}, who introduced the toric code as the first many body \emph{protecting Hamiltonian}.
The promising conjunction of properties shown by his proposal has fueled a search, which is yet to provide a definitive result.

For families of protecting Hamiltonians, such as Kitaev's toric code \cite{kitaev_fault-tolerant_2003, dennis-2002}, a constant energy gap $\gamma$ separates the degenerate ground space, used for encoding, from low energy excited states.
Furthermore, the stabilizer representation of these Hamiltonians naturally associates it with a QECC, which permits an error threshold without the use of concatenation \cite{dennis-2002}.
A perturbation theoretic expansion of local errors $V$ in the Hamiltonian must then cancel to orders increasing with the distance of the associated QECC. 
Thus, within the region of validity of degenerate perturbation theory, the degeneracy of the ground state space is only split by an amount that is exponentially small in the distance of the code. 
In turn, it takes an exponentially long time for the splitting to implement  logical rotations on the perturbed ground space (e.g.~a phase gate).

However, such perturbation theoretic results must be applied with caution.
The most important limitation probably arises from the fact that they deal with a closed quantum system whereas actual noise may be better modeled by perturbative coupling to an environment.
Furthermore, the rigorous range of validity of perturbation theory \cite{kato_perturbation_1995}, $\norm{V} < \gamma/2$, is extremely restrictive when considering perturbations with an extensive operator norm.
Even if local observables can be adapted for to a high degree of accuracy, as shown by Hastings and Wen \cite{hastings}, the global eigenstates of the system may change and become very different. 
Within our understanding, the possibility of adapting encoding and decoding protocols relies on the perturbation being characterized\footnote{A possible exception to this is given by proposals of adiabatic state preparation~\cite{hamma_adiabatic_2008}.}.
However, most of our work will focus on the relevant scenario of an unknown perturbation, where unperturbed (unadapted) encoding and decoding protocols are used. 

Recently, Alicki {\em et.~al.}~have presented results supporting the instability of quantum memories based on Kitaev's 2D toric code \cite{alicki-2008a} and the stability of its 4D version \cite{alicki-2008b} when coupled to a sufficiently cold thermal environment.
Chesi et al. \cite{chesi_thermodynamic_2009} have made progress in providing a general expression giving a lower bound for the lifetime of encoded information.
The approach taken in these articles is thermodynamic in nature and has the advantage of allowing the 
derivation of positive results.
A weak coupling Markovian approximation to an environment at thermal equilibrium is assumed, thus neglecting any memory effects from the environment.
In a previous article \cite{pastawski-2009}, we considered a Hamiltonian system subject to independent depolarizing noise (corresponding to the high temperature limit of the above approach) and proved that $O(\log N)$ is the optimal survival time for a logical qubit stored inside $N$ physical qubits.

Our current approach directly deals with Hamiltonian perturbations and environment couplings without going through a Markovian approximation for the environment.
A comparative advantage of our approach is the capability of exactly dealing with certain weak but finite perturbations and couplings, and providing restricted no go results.

We consider the effect of relatively weak yet unknown perturbations of an $N$ qubit local protecting Hamiltonian and coupling to an ancillary environment starting out in its ground state.
We show that as the number $N$ of physical subsystems used grows, it is impossible to immunize a quantum subspace against such noise by means of local protecting Hamiltonians only.
We further show that if one wishes to recover the quantum state by means of an error correction procedure, the QECC used must have some finite error threshold in order to guarantee a high fidelity; this result is applied to the 3D compass model which is shown not to have such a threshold.
In the case of the 2D toric code, we propose Hamiltonian perturbations capable of destroying encoded information after a time proportional to $\log(N)$, suggesting that some form of macroscopic energy barrier may be necessary.
Weak finite range Hamiltonian perturbations are then presented which destroy classical information encoded into the 2D Ising model; in this case interactions involving a large, yet $N$ independent, number of qubits are required.
Finally, we consider time dependent Hamiltonian perturbations and coupling to an ancillary environment with a high energy density; here we provide constructions illustrating how these more powerful models may easily introduce logical errors in constant time into information protected by all stabilizer Hamiltonians, and certain generalizations.

\subsection{Noise model motivation} \label{sec:StaticPerturbation}

A prerequisite to assess protecting Hamiltonians is a precise definition of the noise model they will be expected to counter.
Our aim is to understand the protection lifetime they provide to (quantum) information as well as to identify the properties a good protecting Hamiltonian should have.
In order to be able to make such predictions, we will study noise models admitting a mathematically tractable description while striving to keep our choices physically motivated.

To falsify claims of protection against {\em any} possible noise of a certain class (such as weak local perturbations to the Hamiltonian), it suffices to consider an adversarial choice  within such a class.
In such a noise model, different perturbations and environments are not assigned probabilities; a perturbation is simply considered possible if it adheres to certain conditions.
Finally, we assume that the perturbation has not been characterized.
This allows us to derive no-go, or limitation, results from the exact analysis of adversarially engineered noise instances.

The most elementary way in which the Hamiltonian evolution of a closed system can be altered is by including a small perturbation $V$ to the Hamiltonian $H$.
A simple physical interpretation for such a perturbation is to associate $V$ to  imperfections in the implementation of the ideal protecting Hamiltonian $H$.
Furthermore, Hamiltonian perturbations extending beyond the system under experimental control are modeled by a weak coupling between the system and an environment.
We focus on families of protecting Hamiltonians satisfying certain locality and boundedness conditions, and naturally extend similar restrictions on the perturbations and couplings considered.

Let us first introduce some definitions.
A family of protecting Hamiltonians $\{H_N\}$ is parametrized by a number $N$ which grows with the number of physical subsystems participating in $H_N$.
A Hamiltonian $H$ is called ``$k$-local'' when it can be represented as a sum
\begin{equation}\label{eq:LocalHamiltonianDecomposition}
 H = \sum_{i=1}^N T_i,
\end{equation}
with at most $k$ physical subsystems participating in each interaction term $T_l$.
The interaction strength of a physical subsystem $s$ in a $k$-local Hamiltonian $H$ is given by the sum $\sum_i \norm{T_i}$ of operator norms over those interaction terms $T_i$ in which the physical subsystem $s$ participates.
A family of $k$-local Hamiltonians is called ``$J$-bounded'' if, for every Hamiltonian $H_N$ in the family, the largest interaction strength among the physical subsystems involved is no greater than $J$.
Finally, a family of Hamiltonians will be D-dimensional if the physical subsystems involved can be arranged into a D-dimensional square lattice, such that all interaction terms are kept geometrically local.

We will concentrate on families of $k$-local, $J$-bounded protecting Hamiltonians, with $J>0$, and $k,J\sim O(1)$. 
Furthermore, the specific Hamiltonians treated in this article admit an embedding into 2, 3 or 4 spatial dimensions and we may assume such embeddings also when dealing with generic protecting Hamiltonians.

The families of Hamiltonian perturbations $\{V_N\}$ which we will consider will be $\tilde{J}$-bounded, with the strength $\tilde{J}$ small in comparison to $J$.
The perturbations will be taken to be $\tilde{k}$-local, with $\tilde{k}$ possibly different, and even larger, than $k$.
This allows, for example, taking into consideration undesired higher order terms which may arise from perturbation theory gadgets \cite{bravyi_quantum_2008}.
Allowed perturbations should also admit a geometrically local interpretation under the same arrangement of subsystems as the protecting Hamiltonian.

When considering coupling to an environment, an additional set of physical subsystems will be included as the environment state.
A family of local environment Hamiltonians $\{H^{(E)}_N\}$ will be defined on these additional  subsystems.
The coupling between system and environment will be given by a family of weak local Hamiltonian perturbations $V^{(SE)}_{N}$, acting on both system and environment.
\begin{equation}
 \tilde{H}_N = H^{(S)}_N \otimes I^{(E)}_N + I^{(S)}_N\otimes H^{(E)}_N + V^{(SE)}_{N} 
\end{equation}
Finally, it should be possible to incorporate the additional physical subsystems from the environment while preserving the number of spatial dimensions required for the Hamiltonian.
To simplify notation, the sub-index $N$ shall in general be dropped.

The engineering of $k$-body interactions is increasingly difficult as $k$ grows \cite{wolf_quantum_2008, bravyi_quantum_2008}.
This is why we limit our study to families of $k$-local Hamiltonians (i.e.~$k$ independent of $N$).
It is under such criteria that we exclude proposals such as quantum concatenated-code Hamiltonians \cite{bacon_2008}, for which the required degree of interactions would grow algebraically with the number of qubits.

The $J$-bounded condition guarantees that the rate of change for local observables remain bounded.
Furthermore, this condition is strictly weaker than the rigorous requirements stated for the simulation through the use of perturbation theory gadgets \cite{bravyi_quantum_2008}.
There, constant bounds are imposed both on the norm of each interaction as well as on the number of interactions in which each subsystem participates.
The $J$-bounded condition also leaves out systems with long range interactions, as, for those systems, the total interaction strength of individual physical subsystems diverges as the system size grows.
Such long range interacting systems are physically relevant, and may lead to protecting Hamiltonian proposals \cite{chesi_self-correcting_2009, hamma_toric-boson_2009}.
However, we abstain from treating such models for which our notion of weak perturbation seems inappropriate.

Each physical subsystem may independently be subject to control imprecision.
Such is the case for weak unaccounted ``magnetic field'' acting on every component of the system or a weak coupling of each component to an independent environment.
Thus, relevant physical scenarios involve perturbations with extensive operator norm (i.e.  scaling with the number of subsystems).
The $\tilde{J}$-bounded condition encapsulates these scenarios and seems to better describe what we understand by a weak perturbation.

Finally, it is expected that scalable physical implementations should be mapped to at most three spatial dimensions.
This would rule out the 4D toric code Hamiltonian \cite{dennis-2002}, a proposal which was otherwise shown to provide increasing protection against weak local coupling to a sufficiently cold thermal bath \cite{alicki-2008b}.
As would occur with an actual physical embedding, we expect that the perturbations considered may be included into the same geometrical picture as the protecting Hamiltonian they affect.

\subsection{Outline of results}

In the following sections, we analyze the problem of obtaining increased protection for quantum information by means of an encoding and a protecting Hamiltonian acting on an increasing number of subsystems.
We consider the effect of adversarial noise models consisting of local Hamiltonian perturbations and/or a weakly coupled environment. 
The aim is to examine the assumptions and limitations of memory schemes based on Hamiltonian protection with a growing number of physical subsystems as quantified by the survival time of stored information.

We prove in complete generality that if the read-out process does not incorporate a recovery procedure then information retrieval will be unreliable.
The figure of merit considered here is $S(t) = \trace{\ket{\psi(0)}\bra{\psi(0)}\rho(t)}$, the overlap between initial and evolved state after a constant time $t$.
For arbitrary protecting Hamiltonians we provide a completely general construction involving a weakly coupled environment starting in its ground state (Sec.~\ref{sec:LowEEnv}) which yields an exponentially small (in $N$) upper bound on $S(t)$ after a constant time.
For gapped Hamiltonians, a proof proceeding without reference to an environment (Appendix \ref{ap:GappedHUnstability}) can provide an upper bound to the time averaged overlap which is close to $\half$.

From this point on, protecting Hamiltonians are considered together with a recovery operation $\R$, applied on read-out, thus providing a more robust figure of merit $S_{\R}(t) = \trace{\rho(0) \R(\rho(t))}$.
A similar weak coupling construction shows that information content of the 3D compass model \cite{bacon-2006} can be destroyed in constant time by a zero temperature environment (Sec.~\ref{sec:bacon}), despite of its error correcting mechanism $\R$.
This provides a direct, negative, answer to the open question of whether the 3D compass model is a self-correcting quantum memory.
From a broader perspective, the structure of our proof strongly suggests that the underlying QECC defining the recovery operation $\R$ must have a strictly positive error threshold.

We continue by considering the effect of Hamiltonian perturbations on the 2D toric code \cite{kitaev_fault-tolerant_2003}.
The recovery mechanism $\R$ is then taken as the composition of a fixed syndrome measurement followed by a correction operation pairing the detected anyons.
It is shown (Sec.~\ref{sec:toric}) that, although the underlying QECC has an error threshold, it is not protected against combinations of {\em unknown} weak local Hamiltonian perturbations, even after a final round of error correction is considered.
Our claim is based on adversarial weak local perturbations that are capable of destroying the stored information in a time logarithmic in $N$.
This is stronger than previous results \cite{kay_2009} in that, the noise model requires no interaction with the environment and the information is destroyed exponentially faster.

In a similar manner, we consider perturbations on the 2D Ising Hamiltonian (Sec.~\ref{sec:ising}), which is often used as an example of self-correcting classical memory.
Here, Hamiltonian perturbations may transform (classical) code states into an ambiguous state in constant time.
While the number $\tilde k$ of bodies in perturbation terms is required to grow as the overall perturbation strength decreases, it shows no dependence on the size $N$ of the system.
In this model, any sequence of local errors connecting the two classical code states must go through states with a macroscopic amount of extra energy, showing that this property alone is not sufficient to give protection.

Beyond the Hamiltonian perturbation model and coupling to a ground state environment, we consider more aggressive noise models (Sec.~\ref{sec:aggressive}) in which the environment can introduce large amounts of energy.
The models considered are time dependent Hamiltonian perturbations and weak Hamiltonian coupling to an environment starting in a high energy state.
For such noise models, even the information storage capabilities of the 4D toric code, a proposal shown to be thermodynamically stable, are completely destroyed.

One might expect that the results presented here are not limited to the task of designing a quantum memory. 
Rather, they tell us about the difficulty of keeping a state and its time evolution confined within a specific subspace of the system, under the effect of Hamiltonian noise. 
Such considerations arise in other settings, such as in the models of adiabatic and topological quantum computation. 
We will outline some of these connections in Sec.~\ref{sec:FurtherApps}.

\section{Necessity of error correction on read-out \label{sec:nec_conds}}
We start our examination of passive quantum memories by proving that a final step of error correction is necessary, regardless of the choice of encoding and protecting Hamiltonian.
What we mean by this, is that if such a recovery step is not included on read-out, there is no way of guaranteeing a high fidelity (close to 1) between initial and final states for more than a constant amount of time.

First, we propose weak local perturbations showing an exponentially decreasing overlap  between perturbed and unperturbed eigenstates.
For general local Hamiltonians and states, we consider a local coupling $V$ of the system with a $\gamma$-bounded environment initialized in its ground state.
Averaging over such couplings $V$, we are able to derive an exponentially small upper bound $\langle S(t_f) \rangle_V \leq [1-\sin^2(2\epsilon)]^N$ for the overlap between initial and evolved system states at a time $t_f=\frac{\pi}{\gamma}$. 

\subsection{Eigenstate susceptibility to perturbations
\label{sec:PerturbedEigenstates}}

Consider a $k$-local Hamiltonian $H$, decomposable into $k$-body interaction terms $T_i$ as described in (\ref{eq:LocalHamiltonianDecomposition}).
We choose a perturbation $V$ such that the initial and final Hamiltonians are related by a composition of local unitary transformations,
\begin{equation} \label{eq:pert}
 \tilde{H} = H + V = \U H \U^\dagger, \textrm{ with }  \U = \bigotimes_{l=1}^N e^{i\epsilon P_l},
\end{equation}
where $P_l$ are normalized local Hermitian operators.
Taking this definition, $V$ can be written as 
\begin{equation}\label{eq:PerturbationDecomposition}
 V = \sum^N_{i=1} \U T_i \U^\dagger - T_i,
\end{equation}
and thus is also $k$-local. 
Furthermore, if $2k\epsilon \ll 1$, it is justified to call $V$ a perturbation with respect to $H$, since all terms are small with respect to those of $H$.

Degeneracies in $H$ are assumed to be infinitesimally lifted to ensure uniquely defined eigenvectors.
The overlap between eigenvectors $\ket{\psi_i}$ of $H$ and the perturbed eigenvectors $\U\ket{\psi_i}$ is then given by $F_{\U}=\abs{\bra{\psi_i}\U \ket{\psi_i}}^2$.

By averaging over all possible directions $P_l$, we effectively obtain an independent qubit depolarization.
\begin{equation}\label{eq:averageasdepolarizing}
 \int \U^\dagger \ket{\psi_i}\bra{\psi_i} \U d P_1 \ldots d P_N = \Delta_{\lambda(\epsilon)}^{\otimes N}(\ket{\psi_i}\bra{\psi_i})
\end{equation}
Here, $\Delta_{\lambda}(\rho)=\lambda \rho + (1-\lambda) \frac{I}{2}$ is the qubit depolarizing channel  and $\lambda(\epsilon) = 1 -\frac{3}{2}\sin^2(\epsilon)$.
We may then denote $\langle F \rangle_{\U}$ as average of the overlap $F_{\U}$ over all local rotations $\U$ having a given strength $\epsilon$.
This average is expressed in terms of the depolarizing channel as
\begin{equation}\label{eq:perturbedeigenstateoverlapbound-1}
  \langle F \rangle_{\U} =  \bra{\psi_i} \Delta_{\lambda(\epsilon)}^{\otimes N}\left(\ket{\psi_i}\bra{\psi_i}\right)\ket{\psi_i}.
\end{equation}
A result of King \cite{king-2003}, known as multiplicativity of the maximum output $p$-norm for the depolarizing channels, states that
\begin{equation}\label{eq:king}
\max_{\ket{\phi}} \norm{\Delta_\lambda^{\otimes N}\left(\ket{\phi}\bra{\phi}\right)}_p \leq
\left( \max_{\ket{\phi}} \norm{\Delta_\lambda \left(\ket{\phi}\bra{\phi}\right)}_p \right)^N .
\end{equation}
For qubit subsystems and for $p=\infty$, Eq. (\ref{eq:king}) bounds the overlap of $\Delta_\lambda^{\otimes N}\left(\ket{\phi}\bra{\phi}\right)$ with any single pure state, leading to
\begin{equation}\label{eq:perturbedeigenstateoverlapbound}
\langle F \rangle_{\U} \leq \left(\frac{1+\lambda}{2}\right)^N = \left(1-\frac{3}{4}\sin^2(\epsilon)\right)^N .
\end{equation}
Not only does this imply the existence of specific rotations such that $F$ becomes exponentially small as the number of subsystems $N$ grows, but that this is true for most rotations $\U$. 
While this is already known under the name of Anderson's orthogonality catastrophe (see, for example \cite{anderson_infrared_1967, paolo_prl99_095701_(2007)}), we re-derive it for completeness and as an  opportunity to introduce techniques needed throughout the rest of the paper.

\subsection{\label{sec:LowEEnv}
State evolution in coupled Hamiltonians}

In this section, we consider a weak Hamiltonian perturbation coupling the system to a ``cold'' environment.
The environment is assumed to start in its ground state, corresponding to a cold environment assumption. 
Averaging over a specific family of such perturbations instances $V$, an exponentially small bound on the overlap between the initial state and the evolved state is obtained.
This bound, $\langle S(t_f) \rangle_V \leq [1-\sin^2(2\epsilon)/3]^N$, is obtained after a constant evolution time $t_f = \frac{\pi}{\gamma}$, inversely proportional to the strength of the environment Hamiltonian.

Suppose that we start with a state $\ket{\psi_0}$ ``protected'' by an $N$ qubit system Hamiltonian $H_S$.
We can introduce a simple environment, composed of $2N$ qubits, each of which starts in its ground state, $\ket{0}$, and which is defined by its Hamiltonian
\begin{eqnarray}
\label{eq:LowEEnvironmentModel}
H &=& H^{(S)} \otimes I^{(E)} + I^{(S)} \otimes H^{(E)}\\
H^{(E)} &=& \gamma\sum_{i=1}^N \ketbra{1+}{1+}^{(E)}_i-\ketbra{00}{00}^{(E)}_i.
\end{eqnarray}
When necessary, we take the supraindices $(S)$, $(E1)$ and $(E2)$ to denote the system, the first, and second components of the environment respectively.
While both environment components will interact with the system, it is the presence of both which will allow a simple interpretation of the induced decoherence as a probabilistic application of local errors.

We again use the trick of considering a perturbed Hamiltonian $\tilde{H}=\U H \U^\dagger$ which results from the weak local rotations $\U=\bigotimes_{j=1}^N U_j$ of the decoupled Hamiltonian $H$.
The rotation elements will involve both system and environment components, $U_j = e^{i\epsilon P^{(S)}_{j} \otimes X^{(E1)}_{j}} $,
where the operators $P^{(S)}_{j}$ are taken to be Pauli-like operators on site $j$ of the system.

The perturbation $V=\U H \U^\dagger - H$ must be decomposable into small local terms.
Such a decomposition for $V$ is given in terms of the decomposition $H^{(S)}=\sum_i T_i$ into at most $k$-body terms.
Each perturbation term $V_i = \U T_i \U^\dagger - T_i$ has an operator norm no greater than $2\epsilon k \norm{T_i}$ and involves up to $2k$-body interactions\footnote{For $\epsilon  k \ll 1 $, a further decomposition of such terms can be provided in which subterms involving $k+b$ bodies are of strength $O(\epsilon^b)$, guaranteeing that the strength of terms decays exponentially with the number of bodies involved.}.
The perturbation required to rotate the environment Hamiltonian terms involve at most 3-body terms and a total norm bounded by $2\epsilon\gamma$.

The initial state, $\ket{\psi_0}\ket{00}^{\otimes N}$ will thus evolve into 
$ e^{-i t \U H \U^\dagger} \ket{\psi_0}\ket{00}^{\otimes N}$.
The survival probability is then $S(t)=\bra{\psi_0} \rho_S(t) \ket{\psi_0}$, where $\rho_S(t) = \trace[E]{\rho(t)}$, and 
\begin{equation}\label{rhot2}
\rho(t)=\U e^{-i t  H } \U^\dagger \ketbra{\psi_0}{\psi_0}\otimes \ketbra{00}{00}^{\otimes N} \U e^{i t H } \U^\dagger.
\end{equation}
Here, $\U$ may be explicitly decomposed as 
\begin{equation}\label{rotationexpansion}
\begin{array}{rcl}
\U &=& \exp( i\epsilon \sum_j P^{(S)}_{j} \otimes X^{(E1)}_{j}) \\
&=& \sum_\vect{p} \cos(\epsilon)^{N-w(\vect{p})} (i\sin(\epsilon))^{w(\vect{p})} P^{(S)}_{\vect{p}}\otimes X^{(E1)}_{\vect{p}} ,
\end{array}
\end{equation}
where $\vect{p}$ denotes a binary vector indicating the sites on which rotations are applied in $P^{(S)}_{\vect{p}}$ and $w(\vect{p})$ is the weight of the bit string $\vect{p}$ (number of non identity factors in $P^{(S)}_{\vect{p}}$).

Now consider a time $t_f = \frac{\pi}{\gamma}$ such that $e^{-i t_f H}$ transforms components of the environment from $\ket{10}$ to $\ket{11}$, while leaving components in state $\ket{00}$ unaltered.
At such a time $t_f$, substituting $\U$ into expressions (\ref{rhot2}) allows explicitly tracing over the environment to yield
\begin{equation}\label{overlap4}
\begin{split}
  \rho_S(t_f)= \sum_{\vect{p}, \vect{q}}
\cos^2(\epsilon)^{2N-w(\vect{p})-w(\vect{q})} 
\sin^2(\epsilon)^{w(\vect{p})+w(\vect{q})} \times\\
 P_{\vect{p}} e^{-i t_f H^{(S)}} P_{\vect{q}} \ketbra{\psi_0}{\psi_0} P_{\vect{q}} e^{i t_f H^{(S)} } P_{\vect{p}}.
\end{split}
\end{equation}
Thus, $\rho_S(t_f)$ may be considered as the density matrix resulting from the independent probabilistic application of the local unitary rotations prescribed by $\U$ on $\ketbra{\psi_0}{\psi_0}$, followed by the evolution under the unperturbed system Hamiltonian, followed by a second round of random application of the local rotations prescribed by $\U$.
Defining
\begin{equation}\label{eq:localError}
 \E_{R,p}(\rho) = p R\rho R + (1-p) \rho,
\end{equation}
and the Hamiltonian evolution
\begin{equation}
 \H_t(\rho) = e^{-i H^{(S)} t} \rho e^{i H^{(S)} t},
\end{equation}
we can take $p=\sin^2(\epsilon)$ and 
define $\rho_{virt} := \left( \bigotimes_{i=1}^N \E_{P_i,p} \right) \ketbra{\psi_0}{\psi_0}$, so that we can express $S(t_f)$ as
\begin{equation}
 S(t_f) = \trace{\rho_{virt} e^{-i t_f H_S} \rho_{virt} e^{i t_f H_S}}. \label{eq:EvolvedSystemState}
\end{equation}
This is the overlap between a density matrix and its own unitary evolution, and can be upper bounded by
\begin{equation}\label{eq:onlyErrors}
 S(t_f) \leq \trace{\rho_{virt}^2}.
\end{equation}
In turn, using the fact that $P_i$ are Pauli-like operators we may rewrite it as
\begin{equation}\label{eq:onlyErrors2}
 S(t_f) \leq \bra{\psi_0} \left( \bigotimes_{i=1}^N \E_{P_i,2p(1-p)} \right) (\ketbra{\psi_0}{\psi_0}) \ket{\psi_0}.
\end{equation}
Averaging over the Pauli-like operators, we obtain 
\begin{equation}
 \langle S(t_f) \rangle_{V} \leq \bra{\psi_0} \Delta^{\otimes N}_{\lambda(\epsilon)} (\ketbra{\psi_0}{\psi_0}) \ket{\psi_0},
\end{equation}
which is the overlap at time $t_f$ averaged over the proposed family of weak perturbative couplings.
Here $\Delta_{\lambda}(\rho)$ is again the depolarizing channel, and  $\lambda(\epsilon)=$ is $1-\frac{4}{3}2p(1-p)$.
Using Eq.~(\ref{eq:perturbedeigenstateoverlapbound}), we obtain
\begin{equation}
 \langle S(t_f) \rangle_{V} \leq [1-\frac{4}{3}p(1-p)]^N ,
\end{equation}
which by substituting $p$ for $\sin^2(\epsilon)$ yields
\begin{equation}
 \langle S(t_f) \rangle_{V} \leq [1-\sin^2(2\epsilon)/3]^N .
\end{equation}
By averaging over different possible weak couplings, we obtain an overlap between initial and evolved states which is exponentially decreasing in $N$.

The norm $\gamma$ of Hamiltonian terms in the environment should be bounded, since it is in part these terms which are rotated by $\U$ to introduce a weak coupling between system and environment.
Thus, the proposed evolution time $t_f = \frac{\pi}{\gamma}$ is constant.
Furthermore, if one considers an environment of $N$ semi-infinite chains of coupled two level systems (such as Heisenberg chains), it is possible to ensure that the overlap with the initial state is small for all times larger than $t \sim \frac{\pi}{\gamma}$, rather than have the recurrences that arise from the discrete spectra of the described model.

In appendix (\ref{ap:GappedHUnstability}), the evolution of an unperturbed eigenstate is considered under the effect of pure Hamiltonian perturbations (no environment).
In this case, a constant rate of change in the system state is guaranteed by an energy gap $\gamma$ in the system Hamiltonian.
If an initial state belongs to an energy band separated from the rest of the Hilbert space by such an energy gap $\gamma$, we provide an upper bound on the time averaged overlap $\langle S(t') \rangle_{t' \in [0,t]} \leq \frac{1}{2}+\frac{1}{2\gamma t}$ between initial and evolved state. 

\subsection{Discussion}

We were able to show after a constant time $t_f$, an exponentially large degradation of the overlap $S(t_f)$ in terms of $N$.
If we are to find a benchmark by which to evaluate a memory scheme, we expect that the memory improves as more resources are dedicated to its implementation.
It is now clear that many-body quantum states are inherently unstable with respect to the uncorrected overlap, i.e. $S(t)$ is not the appropriate benchmark.

The fact that we may count on $N$ physical subsystems to implement a quantum memory should not exclude using only one of them and ignoring whatever noisy evolution is affecting the others.
This corresponds to considering an overlap reduced to the relevant subsystem and not on the whole state.
Already such a simple idea guarantees that information storage quality is non-decreasing with $N$.

One may further generalize this by realizing that the relevant subsystem need not correspond to an actual physical subsystem.
This corresponds to preparing code states of a QECC and performing a single round of error correction $\R$ on read-out.
The resulting benchmark is $S_\R(t)=\trace{\rho(0)\R(\rho(t))}$, quantifying the quality of the information extracted from the evolved state and not of the state itself.
Robust error corrected logical observables\footnote{
Alicki et al. \cite{alicki-2008b} use the name dressed observables, whereas Chesi et al.~\cite{chesi_thermodynamic_2009} use the self-explanatory name of error corrected logical operators, which we will adopt.
} can analogously be defined via the extension provided from the code space to the whole Hilbert space by the recovery operation $\R$.
In the following sections, we shall consider protecting Hamiltonians associated with such error correcting codes and robust logical observables.

\section{Limitations of the 3D compass model \label{sec:bacon}}
A desirable property for a quantum memory is that any sequence of local operators mapping between different logical code-states should have energy penalties which grow with the system size.
It has been shown that this happens for schemes in four dimensions, such as the 4D toric code \cite{dennis-2002}.
Seeking to provide such an example in three spatial dimensions, Bacon proposed the 3D compass model \cite{bacon-2006}, a scheme based on subsystem error correcting codes \cite{kribs_laflamme_poulin_2005} and requiring only 2-body nearest neighbor interactions.
Furthermore, mean field arguments suggest that this model might show such an increasingly large energetic barrier.

However, we will show that the zero temperature (local, but non-Markovian) environment construction of the previous section is capable of giving a false read-out from the code after constant time.
We argue that this is due to the choice of the prescribed error correction protocol by showing that the same flaw is present for the 4D toric code if one uses a similar error correction technique.
We show that this failure is a general feature of all quantum memory protocols for which the underlying quantum error correcting code does not have a local error threshold (is unable to handle errors on a constant fraction of the sites).

 For the 3D compass model, quantum information is first encoded into states of a 2-local Hamiltonian $H_S$ defined on a $N\times N\times N$ arrangement of two level systems (where $N$ is an odd number).
\begin{equation}\label{eq:bacon3DH}
\begin{split}
  H_S = -\lambda\sum_{i,j=1}^N \sum_{l=1}^{N-1} \left( 
X_{l,i,j}X_{l+1,i,j} + X_{i,l,j}X_{i,l+1,j} \right.\\
\left. + Z_{i,l,j}Z_{i,l+1,j} + Z_{i,j,l}Z_{i,j,l+1}
\right).
\end{split}
\end{equation}
This is not a stabilizer code but a subsystem code. 
This means that a recovery operation need not correct certain errors which have no effect on the logical observables, and information may be preserved even if the recovered state is different.

First, note that pairs of planes of operators $\hat Z_l=\prod_{i,j}Z_{i,j,l}Z_{i,j,l+1}$ and $\hat X_l=\prod_{i,j}X_{l,i,j}X_{l+1,i,j}$ commute with the Hamiltonian $H$ for all $l$.  
This also holds for logical operators, which consist of products along a single plane $\bar{Z} \equiv \bar{Z}_l=\prod_{i,j}Z_{i,j,l}$ and $\bar{X} \equiv \bar{X}_l=\prod_{i,j}X_{l,i,j}$ operators respectively, for an arbitrarily chosen $l$.
In the ground space, the choice of $l$ is irrelevant, since the operators $\hat Z_l, \hat X_l$ all have $+1$ eigenvalues.
Provided $N$ is odd, $\bar{X}$ and $\bar{Z}$ anti-commute, giving a qubit algebra. 
In the presence of errors (outside of the ground space), the error corrected logical observables will be defined as the majority vote among plane observables
\begin{equation}
 \bar{Z}_{ec} = \textsf{maj}_l \bar{Z}_l \quad \bar{X}_{ec} = \textsf{maj}_l \bar{X}_l
\end{equation}
where $\textsf{maj}$ stands for a majority vote among the $\pm1$ eigenvalued commuting operators.
Measuring all pairs of adjacent planes $\hat Z_l$ allows a majority vote error correction scheme to be performed on the value of the $\bar{Z}_l$ plane observables without extracting whether the corrected state yields $+1$ or $-1$ values for all such planes.

Considering a perturbation on the system plus an environment, as presented in section \ref{sec:LowEEnv}.
By explicitly developing the final expectation values for the observables of interest ($\trace{\bar{X}_{ec} \rho_S(t_f)}$ and $\trace{\bar{X}_{ec} \rho_S(t_f)}$ ), it can be seen that the information stored in the code will not be reliable after a time $t_f$. 
For this, we can pick up from the evolved state of the system in Eq.~(\ref{eq:EvolvedSystemState})
\begin{equation}\label{eq:}
  \rho_S(t_f) = \left( \bigotimes_{i=1}^N \E_{P_i,p} \right) \circ \H_{t_f} \circ \left( \bigotimes_{i=1}^N \E_{P_i, p} \right) \ketbra{\psi_0}{\psi_0}.
\end{equation}

Since all plane observables ($\bar{X}_l$ and $\bar{Z}_l$) of a given type mutually commute, and also do so with the Hamiltonian, we can independently consider the probability of each plane observable having suffered a flip.
If the $P_i$ in Eq.~(\ref{eq:localError}) are taken to be single $X$ or $Z$ rotations, they will anticommute with overlaping $\bar{Z}_l$ or $\bar{X}_l$ plane observables respectively, changing their value upon an odd number of applications.
Taking the $P_i$ to be simple $Z$ operators, the probability of flipping the value of an $\bar{X}_l$ plane observables by applying $\bigotimes_{i=1}^{N^3} \E_{P_i,p}$ once is given by
\begin{eqnarray}
p_{\text{plane}*} &=& \sum_{i \in odd}^{i\leq N} \cos^{2N^2-2i}(\varepsilon)\sin^{2i}(\varepsilon)\binom{N^2}{i} \nonumber\\
&=& \frac{1-\cos^{N^2}(2\epsilon)}{2},
\end{eqnarray}
which is exponentially close to $1/2$.
Since all observables involved commute with the system Hamiltonian, the probability of observing any result configuration will be preserved by $\H_{t_f}$.
Finally, a second round of errors $\bigotimes_{i=1}^{N^3} \E_{P_i,p}$ will again flip the observed value for each plane with a probability $p_{\text{plane}*}$.
The final independent probability of flipping the value of each plane is
\begin{equation}
 p_{\text{plane}} = 2p_{\text{plane}*}(1-p_{{\text{plane}*}}) = \frac{1- \cos^{2N^2}(2\epsilon)}{2}.
\end{equation}

The proposed correction scheme is equivalent to a majority voting among such planes.
Thus, if more than half the planes suffer such an error, the majority vote will fail.
The probability for incorrectly measuring the error corrected logical observable $\bar{X}_{ec}$ on read-out is then
\begin{equation}
p_{\text{logic}}=\sum_{i=(N+1)/2}^N p_{\text{plane}}^i(1-p_{\text{plane}})^{N-i}\binom{N}{i}.
\end{equation}
Given that $\half[1-\cos^{2N^2}(2\epsilon)] \leq p_{\text{plane}} \leq \half$, we have that
\begin{equation}\label{eq:logicErrorBacon}
\half[1 - N\cos^{2N^2}(2\epsilon)] \leq p_{\text{logic}} \leq \half. 
\end{equation}
Assuming $\epsilon$ to be a small constant independent of $N$, the probability $p_{\text{logical}}$ will exponentially approach $1/2$ for large $N$.
We conclude that the encoding is not robust against the error model posed by local coupling to a cold adversarial environment.

The problem lies in the error correction mechanism rather than the protecting Hamiltonian itself. 
This becomes apparent if one applies a similar analysis to the 4D Toric code. 
There, the suggested noise model does not present a problem, since the usual error correction \cite{dennis-2002} of the 4D toric code has an error threshold.
That is, provided the probability of per-site error, $p=\sin^2\varepsilon$ is below this threshold, there exist error correction criteria which succeed with a probability approaching 1 exponentially with $N$.
On the other hand, we could consider a majority voting version of error correction in this setting, where we measure hyperplanes of $X$ operators, and apply a majority vote to choose the correct result.
In this case, an analysis completely analogous to that of Bacon's 3D compass code would hold proving such a read-out technique unreliable.
We conclude that the error correction procedure of the compass code does not allow sufficient resolution to use any potentially topological properties of the encoded quantum information.

The errors introduced by $\bigotimes_{i=1}^N \E_{P_i,p}$ are sufficiently general to suggest a necessary criterion for Hamiltonian protection of information from weak coupling to a cold environment.
Even if we consider the first round of errors and the Hamiltonian evolution as part of the encoding procedure, information should still be able to withstand the probabilistic application of arbitrary local errors.
This means that the information, either quantum or classical, should be encoded in such a way as to provide a finite error threshold in the thermodynamic (large $N$) limit. Of course, in practice, it will be desirable to have a code with a fault-tolerance threshold such that when we try to implement the round of error correction, faulty operations can be compensated for.

\section{Limitations of the 2D toric code \label{sec:toric}}

We will now show how local Hamiltonian perturbations are capable of introducing uncorrectable errors in Kitaev's 2D toric code Hamiltonian.
The introduction of such errors will strongly rely on the lack of string tension on the toric code, suggesting a macroscopic energy barrier may be a necessary requirement.
A brief introduction to the toric code Hamiltonian is provided in appendix \ref{ap:ToricCode}, and is recommended to the unfamiliar reader.

Logical operations in the 2D toric code can be realized by creating a pair of anyons, propagating them so as to complete a non-trivial loop, and finally annihilating them.
It is roughly such a scheme that will be followed by the perturbations we develop here.
Repeating techniques from section \ref{sec:nec_conds}, we may consider the initial state as containing a superposition of local errors which are interpreted as neighboring anyon pairs.
A perturbation construction due to Kay \cite{kay_2009} allows the deterministic propagation of such anyon pairs along predefined adversarial paths on the lattice.
Syndrome measurement allows restricting to a probabilistic picture where error strings corresponding to anyon propagation paths are present with a predefined probability.
It is finally the recovery procedure which may possibly complete these errors into logical operations by selecting an incorrect anyon matching.

A family of weak local perturbations capable of probabilistically introducing distant anyon pairs will first be presented.
As before, the initial state $\ket{\psi(0)}$ is assumed to be a ground state of the unperturbed Hamiltonian $H$, in this case an $N\times N$ toric code as in Eq. (\ref{eq:ToricHamiltonian}).
After a time $t_f$ proportional to the maximum desired anyon propagation distance $D$, unperturbed syndrome read-out on $\ket{\psi(t_f)}$ will probabilistically detect distant (as well as local) anyon pairs.
Our construction will then be applied to produce a simple set of $O(N)$ distance anyons such that no syndrome based error correction may be reliably applied.
Later, shorter yet more elaborate anyon propagation paths will require explicit analysis of the error correcting probability of different anyon pairing protocols.
In this context, we find weak Hamiltonian perturbations are capable of introducing logical errors with a large probability ($\approx \half$) in a time $t_f$ logarithmic in the system size $N$.

\subsection{Probabilistic introduction of distant anyons}

Kay \cite{kay_2009} showed that local errors (anyons) in the 2D toric code, and other local stabilizer Hamiltonians lacking string tension, can be propagated into logical errors corresponding to almost complete loop operators by a local Hamiltonian perturbation $P$.
While in his work the initial presence of the anyons was assumed,
here, anyons will be introduced with a certain amplitude by a generalization of the Hamiltonian perturbation $P$.

Consider introducing perturbations of the form $V=\U(H+P)\U^\dagger -H$,
where $\U = \bigotimes_i U_i$ decomposes into weak local unitary rotations, and $P$ is, as in \cite{kay_2009}, a weak local perturbation capable of deterministically propagating anyons in a given time $t_f$.
The perturbed Hamiltonian 
\begin{equation}
 \tilde{H}= \U (H+P) \U^\dagger
\end{equation}
induces a time evolution which can be written as
\begin{equation}\label{eq:evolvedTCstate}
\ket{\psi(t)} = e^{-i t \tilde{H}}\ket{\psi(0)} = \U e^{-i t (H+ P)} \U^\dagger\ket{\psi(0)}.
\end{equation}
In this context, $\U$ and $P$ are chosen such that:
\begin{enumerate}
 \item 
Neighboring vertex anyon pairs are created by $\U^\dagger$, with a certain small amplitude $O(\epsilon)$, by applying weak $Z$ rotations on connecting edges.
 \item 
Each of the anyons is deterministically propagated by $P$ along a predefined path.
Thus, local excitation pairs become strings of errors defining new positions for the anyon pair.
 \item 
Finally, $\U$ is unable to remove both the anyons created by $\U^\dagger$ after at least one of them has been propagated.
Moreover, if none were present, $\U$ creates an anyon pair with amplitude $O(\epsilon)$.
\end{enumerate}

Propagation paths for each anyon are not allowed to overlap but are otherwise completely independent.
The propagation of the $i$-th anyon along its path $\ell^{(i)}$ may be attributed to a specific component $P_i$ of $P=\sum_i P_i$.
In turn, each component $P_i$ admits a decomposition 
\begin{equation}
P_i= \sum_{j=1}^{\abs{\ell^{(i)}}} J^{(i)}_j ~T_{\ell^{(i)}_{j-1} , \ell^{(i)}_{j}}, 
\end{equation}
in terms of local interaction terms $T_{p, q}$, where $\ell^{(i)}_j$ are the anyon locations along the path $\ell$.
As in \cite{kay_2009}, the scalar coefficients $J^{(i)}_j$ are chosen to implement a perfect state transfer \cite{christandl-2004,karbach-2005, kay_review_2009} and each term $T_{p, q}$ implement a swap among vertex anyons on $p$ and $q$. 
If $p$ and $q$ are neighboring vertices, $T_{p,q}$ is defined as 
\begin{equation}\label{eq:SimpleAtomicAnyonPropagationTerm}
 T_{p,q} = Z_s 
 frac{\left(\Id -A_{p} A_{q}\right)}{2},
\end{equation}
where $A_{p}$ and $A_{q}$ are the vertex stabilizer operators corresponding to $p$ and $q$ respectively (appendix \ref{ap:ToricCode}) and $Z_s$ is a $Z$ rotation on physical site $s$ corresponding to the edge connecting $p$ and $q$.
Furthermore, by allowing $p$ and $q$ to be next nearest neighbors, it is possible to have  crossing anyon paths $\ell^{(j)}$, $\ell^{(i)}$ without having them overlap in the anyon locations used.
If vertices $p$ and $q$ are not neighbors, the same effect is obtained by substituting $Z_s$ in (\ref{eq:SimpleAtomicAnyonPropagationTerm}) for a tensor product of $Z$ operators along an edge path $\overline{pq}$ from $p$ to $q$,
\begin{equation}\label{eq:AnyonPropagator}
 T_{p,q} = \bigotimes_{s \in ~\overline{pq}} Z_s ~ \frac{\left(\Id -A_{p} A_{q}\right)}{2}.
\end{equation}

The distance $D$ is the maximum number of steps among the different anyon propagation paths $D = \max_i \abs{\ell^{(i)}}$.
It will be taken as $D=N/2-1$ in section (\ref{sec:ToricOrderN}) and as $D=O(\log N)$ in section (\ref{sec:ToricOrderlogN}).
Fixing the strength of perturbation terms in $P$ as $J^{(i)}_j = \frac{\epsilon}{D}\sqrt{j(\abs{\ell^{(i)}}+1-j)}$ allows the perturbation $P$ to remain $\epsilon$-bounded while allowing simultaneous perfect anyon transfer in a time $t_f = \frac{D\pi}{2\epsilon}$.
Similarly to previous sections, by taking the rotations $U_j = e^{i \epsilon Z_j}$ as $\epsilon$ weak, the final perturbation required $V$ will also be composed of $O(\epsilon)$ strength interactions involving at most 8 bodies each.

The quantum state before measurement at time $t_f$ is given in Eq.~(\ref{eq:evolvedTCstate}).
Expanding $\U^\dagger$ from $U_j = e^{i \epsilon Z_j}$, we get
\begin{equation}
 \ket{\psi(t_f)} = \U e^{-i (P+H) t_f} \bigotimes_j (\cos \epsilon \Id_j -i \sin \epsilon Z_j ) \ket{\psi(0)},
\end{equation}
where the index $j$ ranges over sites of non trivial action for $\U$.
The state $\ket{\psi(0)}$ is a ground space eigenstate of $H$, and assuming the locations $j$ on which $\U$ acts are non neighboring, each $Z_j$ will increase the energy respect to $H$ by $\gamma$.
Furthermore since the energy of a state respects to $H$ depends only on anyon number and $P$ is anyon number preserving, we have $[H,P]=0$, allowing us to write
\begin{equation}
 \ket{\psi(t_f)} = \U e^{-i P t_f} \bigotimes_j (\cos \epsilon \Id_j -ie^{-i \gamma t_f} \sin \epsilon Z_j ) \ket{\psi(0)}.
\end{equation}
Since all the propagations in $P$ commute and correspond to exact transfer of each anyon created by $\U^\dagger$ precisely at time $t_f$ , we may write
\begin{equation}
\begin{split}
 \ket{\psi(t_f)} = \U \prod_j (\cos \epsilon \Id -ie^{-i \gamma t_f} \sin \epsilon \bigotimes_{i \in \ell\ell^{(j)}} Z_i ) \ket{\psi(0)}
\end{split}
\end{equation}
where $\ell\ell^{(j)}$ is the path given by the union of $\{j\}$ and the two propagation paths $\ell^{(j+)}$ and $\ell^{(j-)}$ of $P$ corresponding to the each of the two anyons created by $Z_j$.
By expanding $\U$, we obtain
\begin{equation}\label{eq:UnitaryEvolutionExpansion}
\begin{split}
 \ket{\psi(t_f)} = \prod_j \left[
 \cos^2 \epsilon \Id - ie^{-i \gamma t_f} \sin \epsilon \cos \epsilon \bigotimes_{i \in \ell\ell^{(j)}} Z_i  \right. \\
 \left. + i \cos \epsilon \sin \epsilon Z_j + \sin^2 \epsilon e^{-i \gamma t_f } Z_j\bigotimes_{i \in \ell\ell^{(j)}} Z_i
 \right] \ket{\psi(0)}.
\end{split}
\end{equation}

The state $\ket{\psi(t_f)}$ described by Eq.~(\ref{eq:UnitaryEvolutionExpansion}) corresponds to a coherent quantum superposition of applying different error paths.
For such unitary evolutions, initially orthogonal states will remain orthogonal and thus fully distinguishable.
However, there are at least two mechanisms which lead us to consider a mixed density matrix as the final state.
The first, is due to the fact that the actual perturbation applied is not known, and can for instance be taken probabilistically among the family of perturbations described.
The second, is unperturbed syndrome measurement $\M$, which is the first step of a quantum error correction procedure to recover the initial state.

Syndrome measurement $\M$ will probabilistically project the state $\ket{\psi(t_f)}$ into a subspace consistent with a fixed anyon distribution.
This is the first step of the recovery operation $\R=\C\circ \M$, the sequential application of unperturbed syndrome measurement $\M$ followed by a syndrome dependent correction operation $\C$.
Analysis of different correction strategies $\C$ need only focus on the resulting mixed state $\M(\ketbra{\psi(t_f)}{\psi(t_f)})$.
Since for any anyon configuration there is at most one combination of operators yielding it in Eq.~(\ref{eq:UnitaryEvolutionExpansion}), the state $\ketbra{\psi(t_f)}{\psi(t_f)}$ is reduced to a probabilistic application of these operators on $\ketbra{\psi(0)}{\psi(0)}$.
Again, taking $\E_{R,p}(\rho) = p R \rho R + (1-p) \rho$, one may verify that
\begin{equation}
 \M(\ketbra{\psi(t_f)}{\psi(t_f)}) = \bigcirc_j \E_{Z_j, p} \E_{\bigotimes_{i \in \ell\ell^{(j)}} Z_i , p}(\ketbra{\psi(t_f)}{\psi(t_f)})
\end{equation}
with $p=\sin^2 \epsilon$. 
Note that the order of application is arbitrary, since the $\E_{R,p}$ superoperators commute. 
Thus, one may consider independent probabilities $p$ for observing each anyon pair created by $\U^\dagger$ and propagated by $P$ (or unpropagated anyon pairs created by $\U$).
Hence, when instantiating the Hamiltonian perturbation described on a certain set of anyon propagation paths, one need only deal with the independent probabilities of measuring propagated and unpropagated anyon pairs.

\subsection{Simple error loops in $O(N)$ time \label{sec:ToricOrderN}}

The aim of this subsection is to provide a simple ensemble of perturbations, employing the above construction in such a way that resulting anyon configurations are provably ambiguous, by which we mean that a single anyon configuration could have, with equal likelihood, originated from logically inequivalent errors.
This means that for such configurations, the anyon pairing  recovery procedure $\C$ can do no better than guessing, and will complete a logical error with a $50\%$ probability for any possible choice of $\C$.

\begin{figure}[!t]
\begin{center}
\includegraphics[width=0.44\textwidth]{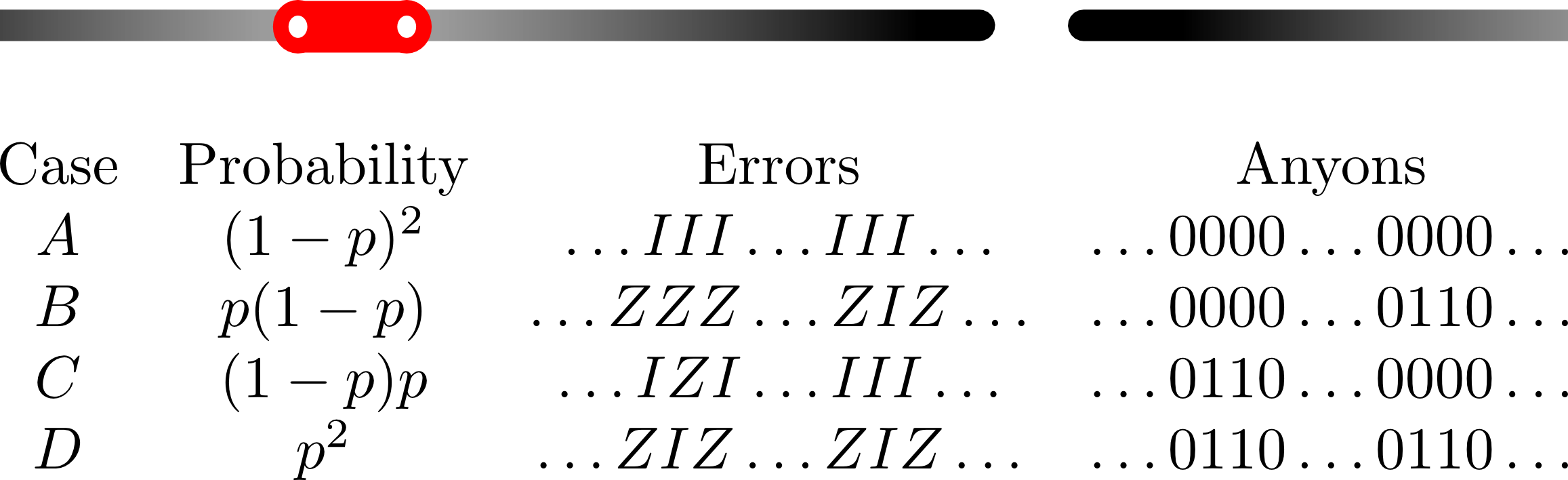}
\end{center}
\caption{
Two hollow dots indicate positions where a pair of vertex anyons may be created by $\U^\dagger$ and/or by $\U$ with probability $p$.
Anyons created by $\U^\dagger$ are propagated by $P$ along the darkening path.
A table is provided indicating the probability of possible error configurations and their corresponding syndrome observables (1 (0) representing anyon presence (absence)). 
\hide{$
\begin{array}{ccc}
\textrm{Probability} & \textrm{Errors} & \textrm{Anyons} \\
(1-p)^2 & \ldots III\ldots III\ldots & \dots 0000 \ldots 0000 \ldots \\ 
p(1-p)   & \ldots ZZZ\ldots ZIZ\ldots & \ldots 0000 \ldots 0110 \ldots \\
(1-p)p   & \ldots IZI\ldots III\ldots & \ldots 0110 \ldots 0000 \ldots \\
p^2      & \ldots ZIZ\ldots ZIZ\ldots  &  \ldots 0110 \ldots 0110 \ldots 
\end{array}
$}
}
\label{fig:LogicalErrors}
\end{figure}

Let us first consider weakly perturbing only in the vicinity of a single row.
The joint effect of many such perturbations, will then be shown to produce further degradation of stored information. So, $\U^\dagger$ introduces a single $Z$ error (neighboring anyon pair) on a physical start site $s$ of the row with probability $p$.
The paths for the perturbation $P$ are chosen such that both of the produced anyons propagate along the row in opposite directions up to final neighboring locations which are  diametrically opposite $s$ (see figure \ref{fig:LogicalErrors}).
For $\epsilon$ weak perturbations, this requires no more than $O(N/\epsilon)$ time.
Finally, with probability $p$, $\U$ may introduce an error at site $s$ or counter an element of the propagated error chain.
As can be seen from the figure, if the anyon introduction site $s$ is chosen uniformly at random, there are observable anyon configurations which occur with probability $2p(1-p)$, which are completely ambiguous (e.g.~cases B and C are indistinguishable under exchange of initial site $s$).
However, if such a syndrome is measured, the correction protocol has a 50\% chance  of completing a horizontal $Z$ loop on the lattice, which is equivalent to applying a completely dephasing channel on one of the encoded qubits with probability $2p-2p^2$.

By applying such a perturbation family to $i\leq N$ rows of the lattice, the probability of not having such a logically dephasing action take place becomes $(1-2p+2p^2)^i$, which may be made arbitrarily small for large $N$ (i.e. an odd number of horizontal $Z$ loops is completed with a probability exponentially close to $1/2$).
Completely analogous string like perturbations exist for any of four logical operators defining the 2-qubit algebra associated to the ground space.
Again, by simultaneously considering such perturbations on a sufficiently large set of parallel lines these operators too will be completed with a probability exponentially close to $1/2$.
Furthermore, by allowing anyons to hop directly to next nearest neighbors (i.e.~Eq.~(\ref{eq:AnyonPropagator})), it becomes possible to simultaneously introduce   perpendicular yet commuting loop operations as a result of anyon removal.

Simultaneously introducing the four logical operators independently with probability exponentially close to $\half$, would yield a state exponentially close to a maximal mixture over the code space.
Our proof requires terms from different perturbation paths to commute, indicating a possible obstacle to achieving this.
In practice however, given that different anyons follow roughly ballistic trajectories with a relatively small spread, this does not pose an issue.
In appendix \ref{ap:ProofOfTCDepolarization}, we show how it is possible to select the set of anyon trajectories in the perturbation such that the order of anyon crossing is well defined (exponentially well in $N$).
In turn, this implies an exponentially small deviation from the result of performing such anyon propagations in order, resulting in a state exponentially close to a maximal mixture on the four dimensional code-space.

\subsection{Localization in 2D stabilizer codes}

In the perturbations constructed to introduce logical errors in the toric code, there is a strong use of the energy degeneracy of subspaces with the same number of anyons.
The strengths of the different stabilizer terms in the 2D toric code manifest as strengths of local magnetic fields in the effective Hamiltonian of the propagation \cite{kay_2009}.
However, having exactly the same strength for all local Hamiltonian terms is not an essential feature of the 2D toric code or of stabilizer Hamiltonians in general.

In the unperturbed picture of stabilizer Hamiltonians, excitations are completely localized.
However, when different excitations live in a degenerate energy space, perturbations may be very effective at propagating them.
With the hope of obtaining some localization with respect to anyon propagation terms, different stabilizer term strengths may be randomly chosen from some range $\gamma_{\text{upper}}>\gamma_{\text{lower}}>0$.

However, a Hamiltonian perturbation may ``smooth'' this distribution to take on, at random, only a finite number of discrete energy values, separated by $\varepsilon$, the strength of the perturbation. 
The number of such possible values is given by 
$\lceil\frac{\gamma_{\text{upper}}-\gamma_{\text{lower}}}{\varepsilon}\rceil$, which is therefore also the average spacing between sites of the same energy. 
Hence, by selecting propagation terms of a similar size, the hopping scheme can route around any defects, and give a logical error.
Thus, the argument is unable to guarantee protection against any constant sized perturbation. 
Nevertheless, it may be that the perturbation terms necessary to break the code should involve a larger number of bodies, which would definitely be an improvement.

In the case of the 2D toric code \cite{kay_2009}, and all other 2D local stabilizer Hamiltonians  \cite{bravyi_terhal_2009, kay-2008b}, there are always logical operations with string like support.
This means that, albeit with some possible energetic smoothening, the scheme presented in section \ref{sec:ToricOrderN} can be adapted to introduce logical errors in arbitrary 2D stabilizer Hamiltonians, meaning that the asymptotic lifetime which 2D $N\times N$ stabilizer codes may guarantee against weak local perturbations cannot be more than $O(N)$.

\subsection{Logical errors in $O(\log(N))$ time\label{sec:ToricOrderlogN}}

\begin{figure}[!t]
\begin{center}
	\includegraphics[width=0.44\textwidth]{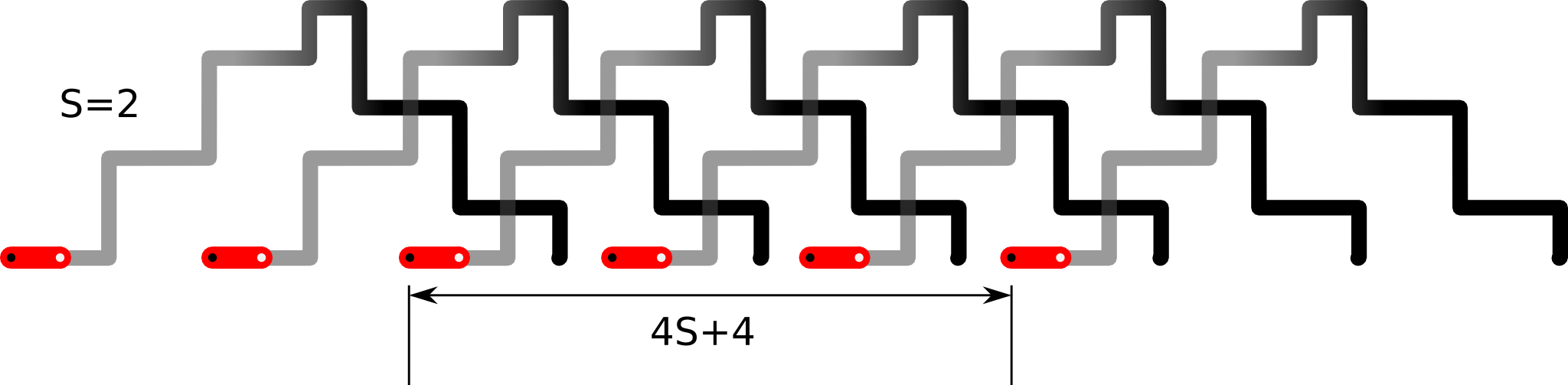}  
\end{center}
\caption{
Anyon pairs corresponding to each thick red edge may be created by $\U^\dagger$.
After a time $t_f$, the right anyon from each pair introduced will be propagated a distance $4S+2$ to the right introducing $Z$ errors along the darkening paths.
Finally, $\U$ acting on the same red segment may move an unpropagated anyon one position to the right or  create a neighboring anyon pair on it.
The number of big steps (or equivalently of crossings) during the upward propagation is given by $S$, which in the case of the figure is 2.
\label{fig:LogicalErrorsFromPerturbation}
}
\end{figure}

In the previous section, we gave a rigorous upper bound of $O(N)$ on the information lifetime of the toric code.
This bound coincides with the one provided by Kay in \cite{kay_2009}, which required initial anyons in the system to be introduced by an unspecified environment.
In this subsection, we provide an exponentially tighter bound by concentrating on specific choices for error correction protocols.
We argue that it is possible for a Hamiltonian perturbation to introduce ambiguous distributions of anyon configurations in a time logarithmic in $N$, i.e.~after a time $t_f \sim O(\log N)$, error correction succeeds with probability not much higher than $1/2$.
Figure \ref{fig:LogicalErrorsFromPerturbation} schematically presents one such perturbation, indicating where anyon pairs should be introduced, and paths $P_k$ along which they should propagate.
The fact that the trajectories have only simple crossings allows them to be implemented by weak local Hamiltonian perturbation terms involving at most 8 bodies, as obtained from  Eq.~(\ref{eq:AnyonPropagator}), with $p, q$ being next nearest neighbors.
Furthermore, the trajectory length is no more than twice the distance at which the anyon pair is finally separated.

The length of anyon propagation trajectories is $8S+4$, where
\begin{equation}\label{eq:define_S}
S=\left\lceil \frac{\ln N}{2p} \right\rceil, 
\end{equation}
and each has $2S$ simple crossings with other trajectories.
The time required to perform such a propagation by fixed strength local perturbations is proportional to $S$ (i.e. logarithmic in $N$).

A relevant property of such a perturbation is that anyons observed when performing unperturbed error correction after an evolution time $t_f$ are always collinear.
The anyon type and line direction may be chosen to coincide with any of the logical operations, translating to the fact that any logical error may be introduced.
This also has the desirable effect of simplifying the analysis of anyon matching criteria. 
There are only two logically inequivalent anyon matchings on the line, which are the two perfect matchings in which each anyon is paired with one of its two nearest neighbors (i.e.~right or left).
The point is that one matching will be logically equivalent to the actual trajectories  performed by the anyons, canceling any errors introduced, whereas the other will complete the actual paths into a logical error.
A simple criterion to determine which case we are dealing with is to count how many times the actual trajectories, together with the anyon matching, cross a vertical line or any homologically equivalent curve.
An odd number of crossings means that a logical error has been completed, whereas an even number of crossings means that the proposed pairing has been successful at error correcting.

We study the success probability of two apparently reasonable matching criteria.
The first minimizes the furthest distance among paired anyons.
The second, for which a polynomial algorithm is known \cite{vaidya_1988}, consists of minimizing the sum of distances among paired anyons.
Proofs and numerics will be provided for the large $N$ regime given by $N \gg 4S+2$ which convey a high logical error rate.

\subsubsection*{ Anyon matching that minimizes $L_\infty$ }

Let us first consider minimizing the furthest distance among paired anyons.
This is the $L_\infty$ norm of the vector with components given by the individual distances among anyons paired by the matching. 
We will prove that the probability of introducing a particular logical error is close to $1/2$ by considering two disjoint scenarios. 
The first is the very unlikely scenario in which, on syndrome measurement, two consecutive anyons are measured at a distance $\geq D$ (by consecutive, we mean no additional anyons were measured in the interval between them).
The second is composed of anyon distributions consistent with the measurement of a fixed pair of consecutive anyons at a distance $\leq D$.
For such distributions, the number of activated anyon paths passing completely over the fixed pair is shown to be odd with probability very close to $1/2$.

Let us first bound the probability of observing two consecutive anyons at a distance greater than $D$ in the syndrome measurement for the evolved state.
Given a fixed region of length $D$, at least $\lfloor D/4 \rfloor$ different potential anyon paths start and end in it.
Furthermore, assuming $D<4S$, the probabilities for not measuring anyons in this region are independent and are $1-p$ for each end of an anyon path and $(1-p)^2$ for each start of an anyon path, since both $\U$ and $\U^\dagger$ could have created anyons in this case.
The anyon-free region can begin in any of $N$ locations of the full loop.
Thus, regardless of correlations, the probability of having $D$ consecutive anyon-free sites is upper bounded by $N(1-p)^{3\lfloor D/4 \rfloor}$.

Assume now that a pair of consecutive anyons is measured at a distance no greater than $D$.
There are at least $\lfloor(4S-D)/4\rfloor$ potential anyon paths going over this region, each with independent probability $p$ of being observed. 
On syndrome read-out, the number of such paths that is activated is odd with a probability approaching $1/2$ at least as fast as $\half(1 \pm (1-2p)^{\lfloor(4S-D)/4\rfloor})$.
Since the $L_\infty$ norm correction completes a logical error if the most distant consecutive anyon pair is covered by an odd number of activated anyon paths, then by inserting $S = \lceil \frac{\ln(N)}{2p} \rceil$ and $D=\lceil \frac{8\ln(N)}{5 p } \rceil$,  we get a probability lower bound for logical errors which approaches $1/2$ as $1/2(1-N^{-1/5})$.

\subsubsection*{ Anyon matching that minimizes $L_1$}

\begin{figure}[!t]
\begin{center}
\includegraphics[width=0.45\textwidth]{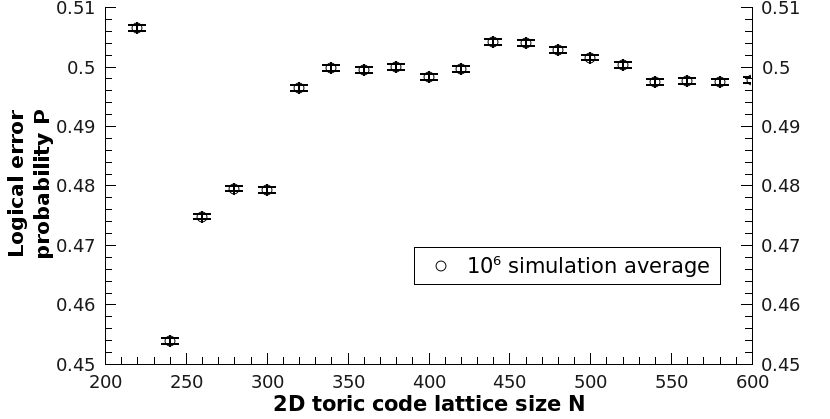}
\end{center}
\caption{
The average probability of error for $L_1$ correction after the system evolves for a time $t_f$ under the described Hamiltonian perturbation.
Here, anyon pairs arise, and evolve to distances of  $\lceil 20\ln(N) \rceil$, with a probability of $10\%$, all collinear on a line of length $N$. 
Each point represents an average over $10^6$ random samples, with error bars representing the magnitude of estimated statistical errors.
}
\label{fig:numerics}
\end{figure}

Let us now consider the anyon pairing criterion that minimizes the total sum of distances among paired anyons.
Since all anyons are found on a loop of length $N$, this criterion will always choose a pairing with total distance no greater than $N/2$.
Thus it will successfully error correct if and only if the total distance of regions of the loop covered an odd number of times by observed anyons is no greater than $N/2$. 
By taking $S$ to be $\lceil \frac{\ln N}{2p} \rceil$, we expect to find roughly half of the sites flipped. 
To see this, note that, on average, each site is covered approximately $S \epsilon$ times.
Moreover, the probability of each site being covered an odd number of times is $\frac{1}{2}[1 -(1-2p)^S]$.
For small $p$ and the chosen value of $S$, the average number of sites covered an odd number of times is approximated by $\frac{N}{2} - \frac{1}{2N}$.
Furthermore, we expect the actual number of such sites to approximately follow a normal distribution around this value, which would imply that logical errors are completed with a probability of close to $\frac{1}{2}$.
However, since the flipping of different nearby sites are highly correlated events, it is not clear how to go about proving this. Instead, computer simulations (Fig.~\ref{fig:numerics}) provide very strong numerical evidence.

\subsection{Discussion}

We have proven that Hamiltonian perturbations can completely destroy the information stored in the 2D toric code in a time proportional to $N$.
The only assumptions are that the precise Hamiltonian perturbation is unknown, and that recovery begins by performing unperturbed syndrome measurements.
A simple family of Hamiltonian perturbations with associated probabilities, was used to justify that the introduction of logical errors in $O(N)$ time is fully independent from the correction protocol used. 
This approach remains applicable for arbitrary 2D stabilizer codes, even when the stabilizer terms are of uneven strength.

Furthermore, we have argued that logical errors may be introduced by weak local perturbations in a time logarithmic with the system size.
In particular, two apparently reasonable anyon pairing schemes were shown to provide an unreliable recovery mechanism against weak local perturbations acting for $O(\log N)$ time.

A fully general proof, including all possible error correction strategies based on syndrome measurement is currently lacking for the $O(\log N)$ error introduction.
The generality of the $O(N)$ construction is obtained by considering a family of different perturbations which could produce the same syndrome outcomes through topologically inequivalent error paths.
It may be fruitful to apply such an approach for a general proof of logical errors produced in $O(\log N)$ time.

\section{Limitations of the 2D Ising model  \label{sec:ising}}
In present day classical computers, magnetic domains are widely used to provide passive safekeeping of classical information.
The Ising model is usually used to elucidate the origin of such long lived magnetized states as a collective effect arising from microscopic local 2-body interactions 
\begin{equation}\label{eq:Ising}
 H_{Ising} = -J \sum_{\langle i, j\rangle}  Z_i Z_j .
\end{equation}
In two and higher spatial dimensions, the nearest neighbor Ising model presents a finite temperature phase transition between a disordered phase and an ordered magnetized phase.
However, it has long been known that such a system looses its asymptotic bistability under the bias produced by even the weakest of magnetic fields \cite{richards_magnetization_1995, cirillo_metastability_1998, grinstein_can_2004}
\begin{equation}
\tilde{H}_{Ising}= -J \sum_{\langle i,j \rangle}  Z_i Z_j + \epsilon\sum_jZ_j .
\end{equation}
Such studies consider the dynamics of minority droplets in a 2D Ising model as given by phenomenological equations or the Metropolis algorithm.
For a Metropolis algorithm in which such a systematic magnetic field $\epsilon$ is included, there is only one stable phase parallel to the field.
The anti-parallel phase becomes metastable, with a lifetime exponential in $J/\epsilon$.
Dependence of the information lifetime on lattice size $N$ appears only for the small $N\lesssim J/\epsilon$ and will thus not appear if one first takes the limit for large $N$.

In this section we consider storing one bit of classical information subject to a quantum evolution of a perturbed 2D Ising Hamiltonian. 
The observable on which the classical bit is encoded is assumed to be the overall direction for magnetization $\bar{Z}=\textsf{maj}_j Z_j$.
The perturbations may conceptually be split into two parts, $Z$ parallel magnetic fields which introduce additional degeneracies to the Hamiltonian, and transverse magnetic fields or many body terms which couple the new ground states, introducing a hopping between them.
The perturbation terms considered will not show support or intensity growing with $N$, and they will be capable of introducing logical errors to the unperturbed logical observable $\bar{Z}$ in a time also independent of $N$.

\subsection{Hamiltonian perturbation proposal}

Consider dividing the $N\times N$ 2D periodic lattice with a chessboard pattern of squares of $M\times M$ spins where $M> 4J/\epsilon_{\text{max}}$ and $J$ is, again, the nearest neighbor Ising coupling constant and $\epsilon_{\text{max}}$ is the greatest local perturbation strength one expects the Hamiltonian to protect information against.
For simplicity, we assume $N = 2nM$, where $n$ is an integer.
Consider alternately introducing $\pm \epsilon Z_j$ magnetic fields in the lattice site $j$ belonging to white/black squares of the chessboard pattern respectively.
The value $\epsilon$ is chosen homogeneously for each square such that the energy difference, $2\epsilon M^2$, from fully field parallel and anti-parallel configurations of each square exactly matches the maximum energy difference for border Ising terms $8MJ$. For $N\gg J/\epsilon$ such a perturbation is always possible. 

The point is that now, the ground space of the system acquires a much higher degeneracy, i.e. between $2^{2n^2+1}-1$ and $2^{4n^2}$.
Each black square could be fully magnetized parallel to its preferred field direction or parallel to the direction of its four neighboring squares if it is opposite.
By taking one spin variable for each square, the ground states may be identified with those of a 2D $n\times n$ anti-ferromagnetic Ising model with magnetic field.
Three important ground states are the two fully magnetized lattice configurations and the checkerboard configuration in which all spins are fully aligned to their local magnetic field.
The gap of these ground states with respect to low lying exited states is $2\epsilon$, which is the energy penalty of flipping a corner lattice site of a square that is fully oriented in the direction of the field but anti-parallel to the two neighboring squares adjacent to the stated corner.

A flipping term for each square of the chessboard should be of the form $\alpha \bigotimes_j X_j$, where the $j$ is taken over all the $M^2$ sites in the square.
Such terms can be introduced either on all black squares or all white squares.
This would respectively couple one of the two fully magnetized configurations with the checkerboard configuration, achieving a full swap of state in a time $t_{\text{flip}} = \frac{\pi}{\alpha}$.
This evolution is exact when such $M^2$-body terms of norm $\alpha$ are allowed, which implies that a proof of Hamiltonian stability will not only require assuming sufficiently weak perturbation terms but also a specific bound for the number of bodies on which such terms act.

Let us now focus on the magnitude of $\alpha$.
This is the coefficient for a many body term, in which the size of the support scales like $M^2 = (4J/\epsilon)^2$, independent of $N$.
One may consider obtaining such a term from the $M^2$-th order degenerate perturbation theory expansion of fields of the form $\epsilon_2 \sum_j X_j$.
For perturbation theory to be strictly valid, one needs $M^2 \epsilon_2 < \epsilon$.
Even then, this small magnitude must be taken to the $M^2$-th power to obtain the first non vanishing expansion term.
The time required to flip all spins in a plaquette is then proportional to:
\begin{equation}
 \alpha \approx \epsilon_2 \left(\frac{\epsilon_2}{\epsilon}\right)^{M^2} \approx  \epsilon M^{-2(M^2-1)} = \epsilon\left(\frac{\epsilon}{4J}\right)^{\frac{32 J^2}{\epsilon^2}+2} .
\end{equation}
This expression has no dependence on $N$ and the same perturbation can be introduced in all squares of a given color to yield a fixed flip time.
Furthermore, we note that the state of those chessboard squares which are not perturbed is fixed and may be traced out exactly.
Hence, the second set of perturbations applied are fully independent and degenerate perturbation theory may be rigorously applied.

\subsection{Discussion}

Although the flip time shows no dependence on $N$, it grows faster than exponentially in terms of $4J/\epsilon$.
It may well be that for magnetic domains, describable by such a 2D Ising Hamiltonian as Eq.~(\ref{eq:Ising}), the ratio $4J/\epsilon$ is sufficiently large to provide a lifetime longer than would be experimentally verifiable.
More importatntly, we are dealing with an extremely simplified model, with the particularity of neglecting any long range interactions of actual physical systems.

The fact that the perturbation is unknown means that if such a checkerboard state is observed on read-out, information is not recoverable.
Such schemes may clearly be generalized to higher dimensions and to deformations of the checkerboard pattern.
The existence of such perturbations elucidates important limitations for statements one may formally prove about the classical memory reliability of the Ising model, and therefore what conclusions one might draw about the presence of a macroscopic energy barrier (string tension) which the 2D Ising model certainly possesses. 
However, it is not clear that these arguments can be applied, for instance, to the 4D toric code since it is a feature of classical memories, but not quantum ones, that local fields can split degeneracies.

\section{Aggressive noise models \label{sec:aggressive}}
In previous sections, we explored the effects of Hamiltonian perturbations on quantum memories, and particularly on the 2D toric code. 
We also considered examples of local cold environments perturbatively coupled to a system, as illustrated by Secs.~\ref{sec:LowEEnv} and \ref{sec:bacon}.
The only energy available in these scenarios was due to local perturbations on the system plus environment. 
Intuitively, a small but constant energy density proportional to $\epsilon$ was allowed.
While this energy is potentially $O(N^d)$ for a $d$ spatial dimension lattice of $N^d$ qubits, it is difficult to concentrate it in specific regions in order to generate logical errors. In comparison, stabilizer codes only require $O(N^{d-1})$ energy to implement a logical gate through local rotations.

More aggressive noise models may locally introduce large amounts of energy into the system while keeping perturbation magnitudes weak.
Such an example is provided by weak yet time dependent Hamiltonian perturbations.
These are relevant when one considers effective protecting Hamiltonians in the interaction picture \cite{brown_energy_2007}.
Another possibility is to consider the weak coupling of the system to an environment which starts in a high energy state.
Noise constructions for these models shall be presented in this section.

In calling such noise models aggressive, we convey the fact that we do not expect ``reasonable'' Hamiltonian protection schemes to guarantee a long lifetime against such models.
Thus, their study may help identify required restrictions on the noise model in order to allow for provably robust Hamiltonian protected memory models.
Furthermore, it may provide insight regarding potentially fruitful proof techniques.

\subsection{Time-varying Perturbations} \label{sec:time_vary}

When considering Hamiltonian perturbations, we assumed that we were unable to determine the new ground space due to the perturbation, and thus encoded in the original ground state space. One might consider an intermediate setting where encoding can be achieved in the perturbed code-space, perhaps due to an adiabatic evolution such as proposed by \cite{hamma_adiabatic_2008}, or by a precise determination of, and compensation for, any perturbations present at the start of the storage time. However, in real experiments, stray fields etc.~responsible for perturbations may fluctuate in time.
Again, if one can track the changes in perturbations, the proof of Hastings and Wen \cite{hastings} continues to hold because Lieb-Robinson bounds apply to time-varying local Hamiltonians, and we can therefore adapt the final error correction step as well. Instead, we proceed assuming it is impossible to precisely learn this time variation. 

\subsubsection{Adiabatically varying perturbations}

One extreme case to consider is that the perturbation varies adiabatically, so that the system remains in its ground state space. 
If we do not apply error correction, then we are concerned with how long it takes before the initial and final ground states have a small overlap. 
We shall assume that the original Hamiltonian $H$ of $N$ qubits has an energy gap $\gamma$, and we will consider the time-varying perturbation
$$
V=\U(t)H \U^\dagger(t)-H
$$ 
where, as before,
$$
\U(t)=\prod_{j=1}^Ne^{-it\epsilon X_j/T}
$$
and $T$ is the total time of the evolution, i.e.~small local rotations are gradually introduced. 
At any time $0\leq t\leq T$, the effective Hamiltonian $\U(t)H\U^\dagger(t)$ has the same energy gap as $H$, which means that the adiabatic condition is satisfied for $T\sim1/\text{poly}(\gamma)$. 
From previous considerations, Eqn.~(\ref{eq:nogapjumpbound}), we know that the overlap of the initial state $\ket{\psi(0)}$ and the evolved state, the ground state of the adiabatically perturbed Hamiltonian, have an average overlap of no more than $\trace{P_0}(1-\frac{3}{4}\sin^2(t\epsilon/T))^N$.
For large $N$ and small $\epsilon$, this means that the final overlap is of the order $\trace{P_0}{\exp}(-\frac{3 \epsilon^2 N}{4})$ if a phase of error correction is not involved.

When error correction is introduced to this scenario, this maps into the situation where our quantum memory is initially encoded in the perturbed subspace, but decoding is using the original, unperturbed, error correction strategy. In the specific instance of the perturbation $\U(T)$, we find that $X$ rotations are applied probabilistically on each site, and hence our QECC must have a superior error threshold.

Hastings and Wen \cite{hastings} reveal a similar interpretation holds for all possible perturbations (assuming the gap remains open as the perturbation is introduced) since all local terms are converted into quasi-local rotations.

\subsubsection{Rapidly oscillating perturbations}

Another extreme scenario is when perturbations are allowed to oscillate with arbitrary frequencies.
A simple construction shows that for stabilizer Hamiltonians, this allows the introduction of arbitrary errors in constant time.
We expect that optimal control theory may provide the tools to generalize such results to arbitrary Hamiltonians.

Consider a stabilizer Hamiltonian $H_0$ and a logical error to implement $L = P_M P_{M-1} \ldots P_2 P_1 $, which is decomposed into Pauli operators $P_i$ on different sites.
We then consider the time dependent Hamiltonian perturbation
\begin{equation}
 V(t) =  \epsilon(t)\sum_i  e^{-i H_0 t} P_i e^{i H_0 t} .
\end{equation}
This perturbation is weak if $\epsilon(t)$ is sufficiently small.
Furthermore, given that $H_0$ is a stabilizer Hamiltonian, $V(t)$ may be written as a sum of local terms (at least as local as the stabilizer operators).
Finally, the dependence of $\epsilon(t)$ on time, is to allow for $\epsilon(0)=0$ which makes the initial encoding equivalent for both perturbed and unperturbed Hamiltonians.

The point of such a perturbation, is that it is possible to explicitly calculate the evolution of the system state in the interaction picture.
\begin{equation}
 \ket{\psi_{I}(t)} = \Pi_i e^{-i P_i \int_0^t \epsilon(t') dt'} \ket{\psi_{I}(0)}
\end{equation}
This means that after a constant time $t_f$ such that $ \frac{\pi}{2} = \int_0^{t_f}  \epsilon(t') dt'$, the target operation $L$ is perfectly implemented in the interaction picture.
If $\ket{\psi(0)}$ is an eigenstate, then $L$ is also implemented in the Schr\"odinger picture, modulo a global phase.

Taking $L$ to be a logical operator of the stabilizer code used, this means that time dependent perturbations of sufficiently high frequency can destroy stored information in constant time.
Here, sufficiently high frequency refers to having perturbation terms which oscillate with frequencies at least as high as those corresponding to localized excitations.

\subsection{\label{sec:HighEEnv}
Stabilizer Hamiltonians and energetic environment}

In what follows, we consider a model in which an environment starts out in an arbitrarily energetic state.
However, the couplings between system and environment are required to remain small and local.

For simplicity, we assume that the system is defined by a stabilizer Hamiltonian $H_S$ and that it starts out in an eigenstate $\ket{\psi_0}$ of all stabilizer operators.
We will consider a sequence of $M$ Pauli operators on different sites $L = P_M P_{M-1} \ldots P_2 P_1 $ compounding to a logical operation.
In the case of translationally invariant stabilizer codes, explicit constructions for these operators are given in \cite{kay-2008b}.
Finally, we may assume a code state $\ket{\psi_0}$, such that $\bra{\psi_0}L\ket{\psi_0}=0$.

Motivated by the realization that, in order to introduce logical errors, we need to transfer some energy from the environment to the system, we choose a specific environment Hamiltonian $H_E = -H^*_S$ (at this point, the complex conjugate is unnecessary, but will become useful later). This means that all steps up in energy in the system correspond to an identical step down in energy in the environment. We start the environment state in $\ket{\psi^*_0}$.

In this scenario, the coupling 
\begin{equation}
 H_{SE} = \epsilon \sum_{i=1}^M P_{S,i}\otimes P_{E,i}^*
\end{equation}
is enough to produce the logical error $L$ in constant time $\frac{\pi}{2\epsilon}$. 
To see this, consider the two states $P_{S,\vect{i}}\ket{\psi_0}P_{E,\vect{i}}^*\ket{\psi_0^*}$ and $P_{S,\vect{i}}P_{S,i}\ket{\psi_0}P_{E,\vect{i}}^*P_{E,i}^*\ket{\psi_0^*}$.
Here, the subindex $\vect{i}$ is an arbitrary binary vector indicating which values of $j$ a product of $P_{S,j}$ (respectively $P^*_{E,j}$) should be taken over.
First of all, since we have assumed $H_S$ is a stabilizer, and the $P_{S,i}$ are Pauli operators, the aforementioned states are zero eigenstates of $H_S\otimes \Id_E + \Id_S\otimes H_E$.
Furthermore, the effective Hamiltonian for the perturbation term $\varepsilon P_{S,i}\otimes P_{E,i}^*$ acting on the pair of states $P_{S,\vect{i}}\ket{\psi_0}P_{E,\vect{i}}^*\ket{\psi_0^*}$ and $P_{S,\vect{i}}P_{S,i}\ket{\psi_0}P_{E,\vect{i}}^*P_{E,i}^*\ket{\psi_0^*}$ is just a matrix
$$
\varepsilon
\left(
\begin{array}{cc}
0 & 1 \\
1 & 0
\end{array}
\right),
$$
independent of $\vect{i}$. This means that we can consider the action of the different $P_{S,i}\otimes P_{E,i}^*$ terms independently:
$$
e^{-iH_{SE}t}\ket{\psi_0}=\left(\bigotimes_{i=1}^{M}e^{-i\varepsilon P_{S,i}\otimes P_{E,i}^*t}\right)\ket{\psi_0}.
$$
Due to the effective Hamiltonian, $P_{S,\vect{i}}\ket{\psi_0}P_{E,\vect{i}}^*\ket{\psi_0^*}$ is mapped to $P_{S,\vect{i}}P_{S,i}\ket{\psi_0}P_{E,\vect{i}}^*P_{E,i}^*\ket{\psi_0^*}$ in a time $\pi/(2\varepsilon)$. Thus, the effect of the entire perturbation is to rotate, in a time $\pi/(2\varepsilon)$ from $\ket{\psi_0}$ to $L\ket{\psi_0}$.

Of course, this approach requires the environment state to have a very high initial energy, namely to start in one of its highest energy states. A refinement of this argument allows us to only change the sign of stabilizers in $H_E$ which share support with $L$. For a local stabilizer code in $d$ spatial dimensions consisting of $N^d$ qubits, it was shown \cite{kay-2008b, bravyi_terhal_2009} that there are logical operators $L$ with support on $k \propto N^{d-1}$ sites.
The initial state of the environment $\ket{\psi_E}=\ket{\psi_0^*}$ is still an eigenvector of these stabilizers, with the same eigenvalues. 
Thus, coupling to an environment with an energy proportional to $N^{d-1}$, may also introduce logical errors in the same time.
This means that the energy required from the environment per system qubit tends to $0$ as $\frac{1}{N}$ (compare to perturbations, which introduce an energy $\varepsilon$ per site).
The catch however, is that the distribution of the energy in the initial environment is highly specific and is in general very different from distributions that may be provided by low temperature thermal states.

We conclude that no stabilizer Hamiltonian will be capable of providing a guarantee for the logical integrity of stored information under the presence of an adversarial, weakly coupled, local environment.
Further statistical assumptions such as energy distribution associated to a low temperature environment state need to be included in addition to the weak local coupling assumptions.

\subsection{Non-stabilizer Hamiltonians}

Stabilizer Hamiltonians are not the only possible candidates for providing information protection, although they are particularly attractive because local errors remain as local errors (not propagating or multiplying in the absence of perturbations). 
Let us now consider the more general case of distance preserving Hamiltonians i.e.~ones which might not leave local errors perfectly localized, but do not increase the distance of the error as defined by an error correcting code\footnote{As an aside, note that Bacon's 3D compass code \cite{bacon-2006} is an example of a code where the errors do not remain in fixed positions, but preserve the values of the observables (since the observables commute with the Hamiltonian), which reminds us (Sec.~\ref{sec:bacon}) that error correction is already likely to become much more problematic for these codes.}.
Using the same construction as in the previous section, we will show that weak coupling to an environment can also introduce the relevant logical error into a distance preserving Hamiltonian (i.e.~a logical operation converting between the most distant code states) for classical memories, by which we mean that one set of local errors becomes irrelevant, say $Z$ errors, and the presence of a logical error on the classical bit depends only on the local $X$ errors present. 
The distance preserving assumption means that the number of $X$ errors is preserved, $[H_S, \sum_i Z_i]=0$. The maximum distance between any 2 states is for the eigenstates $\bigotimes_i \ket{0}_i$ and $\bigotimes_i \ket{1}_i$, suggesting we should use these states for encoding.

Similarly to the previous subsection, we introduce an environment, and a perturbative coupling between system and environment,
\begin{equation}
 H(\epsilon) = H_S \otimes \Id_E + \Id_S \otimes H_E + \epsilon\sum_i X_{i,S} \otimes X_{i,E} .
\end{equation}
The system Hamiltonian is weakly coupled to a ``mirror'' system $H_E = - H_S^*$.
This perturbative coupling is responsible for the evolution of a mirrored state $\ket{0}^{\otimes N}\ket{0}^{\otimes N}$, eigenstate of the unperturbed Hamiltonian $H(0)$.

To analyze the evolution, let us consider the action of the operators $X_i$ in terms of the eigenstates of $H_S$.
Due to the commutation relation, there must be $\binom{N}{m}$ eigenstates $\{\ket{\psi_{m,j}}\}$ of $H_S$ which are simultaneous eigenvectors of $\sum_i Z_i$ with eigenvalue $2m-N$.
We can thus express the eigenvectors $\ket{\psi_{m,j}}$ of $H_S$ in terms of the canonical basis as
\begin{equation}
 \ket{\psi_{m,j}} = 
\sum_{\vect{i}: w(\vect{i})=m} \alpha(m)_{\vect{i},j} X_{\vect{i}}\ket{0}^{\otimes N}  = 
\sum_{\vect{i}: w(\vect{i})=m} \alpha(m)_{\vect{i},j} \ket{\vect{i}},
\end{equation}
where $\vect{i}$ are binary vectors with $m$ non zero components and $\ket{\vect{i}}$ are the respective states from the canonical basis.
The matrix $\alpha(m)$ is unitary as it relates two orthonormal bases of the same subspace.
Define
\begin{eqnarray}
 \ket{\overline{m}} 
&=& \frac{1}{\sqrt{\binom{N}{m}}} \sum_{\vect{i}: w(\vect{i})=m} \ket{\vect{i}} \ket{\vect{i}} \nonumber\\
&=& \frac{1}{\sqrt{\binom{N}{m}}} \sum_{j,k,\vect(i): w(\vect{i})=m}  \alpha(m)^*_{\vect{i},j} \alpha(m)_{\vect{i},k} \ket{\psi_{m,j}}\ket{\psi^*_{m,k}} \nonumber\\
&=& \frac{1}{\sqrt{\binom{N}{m}}} \sum_j \ket{\psi_{m,j}}\ket{\psi^*_{m,j}}.
\end{eqnarray}
From this, one obtains that
\begin{equation}
 \left(H_S\otimes \Id_E - \Id_S \otimes H_S^*\right) \ket{\overline{m}} = 0 , 
\end{equation}
implying that any non trivial evolution of $\ket{\overline{m}}$ arises exclusively from the perturbative coupling and is given by
\begin{equation}
 H(\epsilon)\ket{\overline{m}} = \epsilon J_{m} \ket{\overline{m-1}} + \epsilon J_{m+1} \ket{\overline{m+1}},
\end{equation}
with $J_m= \sqrt{m(N+1-m)}$.
These are precisely the coefficients performing perfect state transfer between $\ket{\overline{0}}=\ket{0}^{\otimes 2N}$ and $\ket{\overline{N}}=\ket{1}^{\otimes 2N}$ in a constant time $t=\frac{\pi}{2\epsilon}$ \cite{christandl-2004, kay_review_2009}.

These results exclude the possibility of proving robustness against weak adversarial coupling to an arbitrarily initialized environment, even of many classical memories using the repetition code (such as Ising models).
We learn that if the environment can provide enough energy, then even weak local couplings may be sufficient to produce logical operations. This also motivates the desire to encode in the ground state space of the Hamiltonian since, were we to encode in a higher energy subspace, the environment needs less energy to cause destructive effects. Alternatively, the mechanism presented here could present a useful way to implement gates on a memory.

\section{\label{sec:FurtherApps}
Further applications}
Constructed perturbations and results presented in this article have focused on elucidating limitations of passive quantum memories.
However, our results may be recast in the following other scenarios.

\textbf{Adiabatic Quantum Computation-}
The standard approach to adiabatic quantum computation consists of implementing an adiabatic evolution 
\begin{equation}
H(t)=f(t)H_i+(1-f(t))H_f, 
\end{equation}
between Hamiltonians $H_i$ and $H_f$, where $f(0)=1$ and $f(T)=0$.
While the ground state of the initial Hamiltonian $H_i$ is expected to be readily prepared, the ground state of the final Hamiltonian $H_f$ encodes the result of the desired quantum computation.
An energy gap no less than $\gamma$ between ground state and excited states of $H(t)$ is required for the duration of the adiabatic evolution.

In this context, it is possible that Hamiltonian perturbations could change the initial or final ground state, and maybe even close the gap during the Hamiltonian trajectory.
For example, a time dependent perturbation
\begin{equation}
V(t)= \left(\U H(t) \U^\dagger-H(t) \right),
\end{equation}
with $\U$ defined as in Eq.~(\ref{eq:pert}), can make the perturbed initial and final ground states almost orthogonal to the unperturbed versions (see Eq.~(\ref{eq:perturbedeigenstateoverlapbound})), while keeping the same gap as $H(t)$.
Even assuming the perturbed initial ground state is exactly prepared, only if the final state belongs to a code space with an error threshold, will it be possible to reliably recover the desired result, as in \ref{sec:time_vary}.

Connections between adiabatic quantum computation and passive quantum memories can be expected to continue into the regime where error correction is incorporated, and future studies may better elucidate the issues involved in developing a fault-tolerant theory of adiabatic quantum computation \cite{lidar-2008}.

\textbf{Topological Quantum Computation-}
Difficulties in implementing quantum memories can also be related to some of the difficulties in implementing a topological quantum computation. 
In particular, in section \ref{sec:toric} we illustrated how constant Hamiltonian perturbations can create and propagate anyons in the 2D toric code.
In the context of topological quantum computation, where gates are implemented through the braiding of anyons, the existence of perturbations capable of creating and propagating anyon pairs is at least equally disturbing as in the memory scenario. 

\textbf{Quantum Simulations-}
 One of the most interesting uses of a quantum computer is likely to be the simulation of other quantum systems. 
While one could express these simulations in terms of the circuit model of quantum computation, and from there create a circuit-based theory of fault-tolerance for quantum simulation, it would be advantageous to understand how this could be implemented more directly, via the simulation of an encoded Hamiltonian. 

A logical first step would be to encode the state of each subsystem to be simulated into a quantum memory. 
Thus, establishing when quantum memories exist, or when they fail, has implications in this case. 
One of the most commonly applied techniques in Hamiltonian simulation is that of the Trotter-Suzuki decomposition, where pulses of non-commuting Hamiltonians are combined into one effective Hamiltonian to some accuracy $\delta$.
This inaccuracy may be treated as a time dependent Hamiltonian perturbation.
Given the power such perturbations were shown to have, it is with great care that one should consider the use of passive quantum memories as elements for such quantum simulators.

\section{
\label{sec:Conclusions}
Conclusions}
In this paper, we have studied several constraints on the extent to which a many body Hamiltonian can be expected to protect quantum information.
First of all, we showed that gapped local Hamiltonians have eigenstates which are asymptotically unstable under local Hamiltonian perturbations. 
This result, commonly referred to as Anderson's orthogonality catastrophe \cite{anderson_infrared_1967} shows that a gap is not sufficient to guarantee protection against errors \cite{jordan_farhi_shor_2006, hamma_toric-boson_2009}.
We proved that a weakly coupled cold environments, can in constant time, alter the evolution of any quantum state leading to an exponentially small overlap between initial and final states.
Taking these results together, we conclude that quantum memory schemes must incorporate  at least one final round of error correction. 

Similar perturbations show why the 3D compass model \cite{bacon-2006}, a self-correcting quantum memory proposal, is not reliable, revealing that the code and error correcting process must posses an error correcting threshold.
Similar conclusions may be drawn in scenarios where information encoding and evolution is according to a perturbed Hamiltonian but read-out and decoding are not.

Further explicit counterexamples illustrate that while we may, in some cases, be able to adapt for known perturbations, arbitrary unknown perturbations can destroy the storage properties of codes such as the 2D Toric code in a time $O(\log N)$.
In this case, the proposed adversarial Hamiltonian perturbation heavily relies on the absence of a macroscopic energy barrier (it is possible to transform orthogonal encoded states via a sequence of local operations while keeping intermediate states in a low energy subspace). By considering the 2D Ising model, we have argued that, in and of itself, a macroscopic energy barrier is not sufficient to protect against perturbations.

Finally, we have considered strong noise models such as time varying Hamiltonian perturbations and weak coupling to an arbitrarily initialized environment.
We showed that these noise models could apply logical transformations on information protected by stabilizer Hamiltonians or distance preserving classical memories in constant time.
We expect these result to provide insight into how one may prove properties of passive quantum memories and under which assumptions.
For instance, since such time-varying Hamiltonian perturbations can destroy the 4D toric code, then when trying to prove robustness against static perturbations, Lieb-Robinson bounds are unlikely to be beneficial.

Having proven a variety of limitations for quantum memory models and elucidated some required conditions, the next step is to incorporate this deeper understanding into new designs for quantum memories. 
One major route is to establish a set of necessary and sufficient conditions under which a quantum memory is protected against unknown weak static perturbations.
Under such a model, we may once again raise the question of whether good protecting Hamiltonians in two or three spatial dimensions exist.
Furthermore, one would hope to find similar conditions under an extended perturbation model allowing a perturbatively coupled local environment.
Here, a central problem is to determine which physically realistic assumption may be made on the environment such that positive results are still attainable (i.e.~conditions on the initial state of the environment, such as it being prepared in its ground state).
Finally, one may study the possibility of engineering an out of equilibrium environment to provide additional protection to quantum information.

We acknowledge the DFG for support (projects Munich Advanced
Photonics and FOR635), and the EU (QUOVADIS and SCALA).
We would like to thank the hospitality of the Pedro Pascual center for sciences at Benasque while part of this work was being done.
ASK is also supported by Clare College, Cambridge.
FP thanks Miguel Aguado for helpful discussions.

\appendix{}

\section{State evolution in perturbed gapped Hamiltonians
\label{ap:GappedHUnstability}}
Energy gaps are considered as a positive feature for a protecting Hamiltonian, since they are expected to provide an energetic barrier which an error process is required to overcome.
However, it will be shown that for sufficiently large $N$, the fidelity of the unperturbed eigenstates acquires an upper bound close to $1/2$ after being evolved under the effect of a perturbed Hamiltonian for a time inversely proportional to the gap energy $\gamma$.

If a system is perturbed, but we do not know the nature of the perturbation, the best strategy is, arguably, to continue using the unperturbed encoding (i.e. the eigenstates of the unperturbed Hamiltonian). 
The survival probability for an unperturbed eigenstate $\ket{\psi_0}$ of $H$ after evolution under a perturbed Hamiltonian $\tilde{H}$ for a given time $t$ (Eq.~(\ref{eq:pert})) is, without error correction
\begin{equation}\label{eq:overlap}
 S(t) = \abs{\bra{\psi_0} e^{-i t \tilde{H}} \ket{\psi_0}}^2 = \abs{\bra{\psi_0} \U e^{-i t H} \U^\dagger \ket{\psi_0}}^2 ,
\end{equation}
i.e. we can express $S(t)$ as the overlap of $\U^\dagger \ket{\psi_0}$ with itself under the evolution of the unperturbed Hamiltonian $H$. 
Furthermore, in terms of the eigenstate decomposition
\begin{equation}\label{eq:eigenstatedecomposition}
 \U^\dagger \ket{\psi_0} = \sum_{j} \alpha_j \ket{\psi_j} \quad \text{ where }\quad H\ket{\psi_j}=E_j\ket{\psi_j},
\end{equation}
$S(t)$ may be expanded as
\begin{equation}\label{eq:overlap2}
 S(t) = \sum_{i,j} \abs{\alpha_i}^2 \abs{\alpha_j}^2 \cos[(E_i-E_j)t].
\end{equation}

\begin{figure}
\begin{center}
 \includegraphics[width=0.15\textwidth]{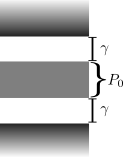}
\end{center}
\caption{\label{gappedSpectrum}There is an energy gap $\gamma$ separating the eigenenergies corresponding to an exponentially small subspace $P_0$ from the energies of the Hamiltonian eigenstates giving rise to the rest of the Hilbert space.}
\end{figure}

Assume the initial state $\ket{\psi_0}$ belongs to an energy subspace $P_0$ of $H$ (i.e. $\bra{\psi_0}P_0\ket{\psi_0}=1$), and that $H$ imposes an energetic gap $\gamma$ between the subspace $P_0$ and its orthogonal subspace (see figure \ref{gappedSpectrum}).
This allows the sum in Eq.~(\ref{eq:overlap2}) to be split as
\begin{equation}\label{eq:overlap3}
\begin{split} 
 S(t) &=
 \sum_{\ket{\psi_i},\ket{\psi_j} \in P_0} \abs{\alpha_i}^2 \abs{\alpha_j}^2 \cos((E_i-E_j)t) \\
 & + \sum_{\ket{\psi_i},\ket{\psi_j} \not\in P_0} \abs{\alpha_i}^2 \abs{\alpha_j}^2 \cos((E_i-E_j)t) \\
 &+ 2\hspace{-2.6ex}\sum_{\ket{\psi_i} \in P_0 ,\ket{\psi_j} \not\in P_0}      \hspace{-2ex}\abs{\alpha_i}^2 \abs{\alpha_j}^2 \cos((E_i-E_j)t) 
\end{split}
\end{equation}
We then define $R$, the $\U P_0 \U^\dagger$ subspace overlap of $\ket{\psi_0}$ as
\begin{equation}
 R =\bra{\psi_0}\U P_0 \U^\dagger \ket{\psi_0} = \sum_{\ket{\psi_i} \in P_0} \abs{\alpha_i}^2.
\end{equation}
Taking the time average $\langle {S}(t') \rangle_{t'\in [0,t]} = \frac{1}{t}\int_0^t S(t') dt'$, and noting that 
\begin{equation}
 \abs{E_i-E_j}\geq \gamma ~\Rightarrow~ \abs{\int_0^t \cos\left((E_i -E_j) t'\right) dt'} \leq \frac{1}{\gamma} ,
\end{equation}
Poincar\'{e} recurrences are averaged out, providing a bound
\begin{equation}\label{eq:overlap4}
 \langle S(t') \rangle_{t'\in [0,t]} \leq R^2 + (1-R)^2 + \frac{2}{\gamma t} R(1-R).
\end{equation}
Although the bound in Eq.~(\ref{eq:overlap4}) is minimized for $R=1/2$, this does not imply that the smallest values for $\langle S(t') \rangle_{t' \in [0,t]}$ are actually obtained for $R=1/2$.

A sufficient condition for the existence of a weak perturbation yielding $R=\frac{1}{2}$ may now be obtained by means of continuity arguments.
First, note that $R$ depends continuously on the parameter $\epsilon$ appearing in the definition of the rotation $\U$, and $R=1$ for $\epsilon=0$.
This means that if, for some $\epsilon_0>0$, we find that $R<1/2$, then $R$ must be equal to $1/2$ for some smaller positive value $0<\epsilon<\epsilon_0$.

As in the previous subsection, we may take $\langle R \rangle_{\U}$ as an average of the overlap $R $ over different directions of the rotation $\U$.
An expression for $\langle R \rangle_{\U}$, in terms of the depolarizing channel is given by
\begin{equation}
\langle R \rangle_{\U} = \trace{ 
P_0 \Delta_{\lambda(\epsilon)}^{\otimes N}\left(\ket{\psi_i}\bra{\psi_i}\right) }.
\end{equation}
Including the dimension of the subspace $P_0$, the same bound as in Eq. (\ref{eq:perturbedeigenstateoverlapbound}) may be used, leading to 
\begin{equation}\label{eq:nogapjumpbound}
 \langle R \rangle_{\U} \leq \trace{P_0}\left(1-\frac{3}{4}\sin^2(\epsilon)\right)^N .
\end{equation}

If the asymptotic growth of $\trace{P_0}$ is slower than $\left(1-\frac{3}{4}\sin^2(\epsilon)\right)^{-N}$, the bound (\ref{eq:nogapjumpbound})
will be exponentially decreasing with $N$.
This means that for sufficiently large $N$, and for most directions of rotation, there is some small rotation parameter $\epsilon$ yielding $R=1/2$.
For the important case of small $\epsilon$ and a constant dimension $\trace{P_0}$, large $N$ refers to $N\sim O(\epsilon^{-2})$.

For those $\U$ leading to $R=\frac{1}{2}$, the time averaged survival probability $\langle S(t') \rangle_{t' \in [0,t]}$ for the corresponding perturbation may be bounded as
\begin{equation}\label{eq:overlap5}
 \langle S(t') \rangle_{t' \in [0,t]} \leq \half + \frac{1}{2 \gamma t}.
\end{equation}
We thus obtain that the overlap of initial encoded states and uncorrected evolved states will drop to values not much larger than $\frac{1}{2}$ in a time inversely proportional to the gap $\gamma$.

\section{\label{ap:ToricCode}
The toric code}
Kitaev introduced the toric code \cite{kitaev_fault-tolerant_2003} with the intention of achieving reliable storage of quantum information at the physical level, as in classical stable storage, rather than by periodically performing explicit error correction procedures.
He proposed that the Hamiltonian of the physical system being used to store the quantum information could, by its nature, make the information stable.
His proposal consisted of a 2D system with non trivial topology (such as the surface of a torus) with a stabilizer Hamiltonian composed of local terms.
Qubits could then be stored in the ground subspace with a degeneracy of $4^g$, with $g$ being the genus of the surface on which the physical qubits are located.

\begin{figure}
\begin{center}
  \includegraphics[width=0.22\textwidth]{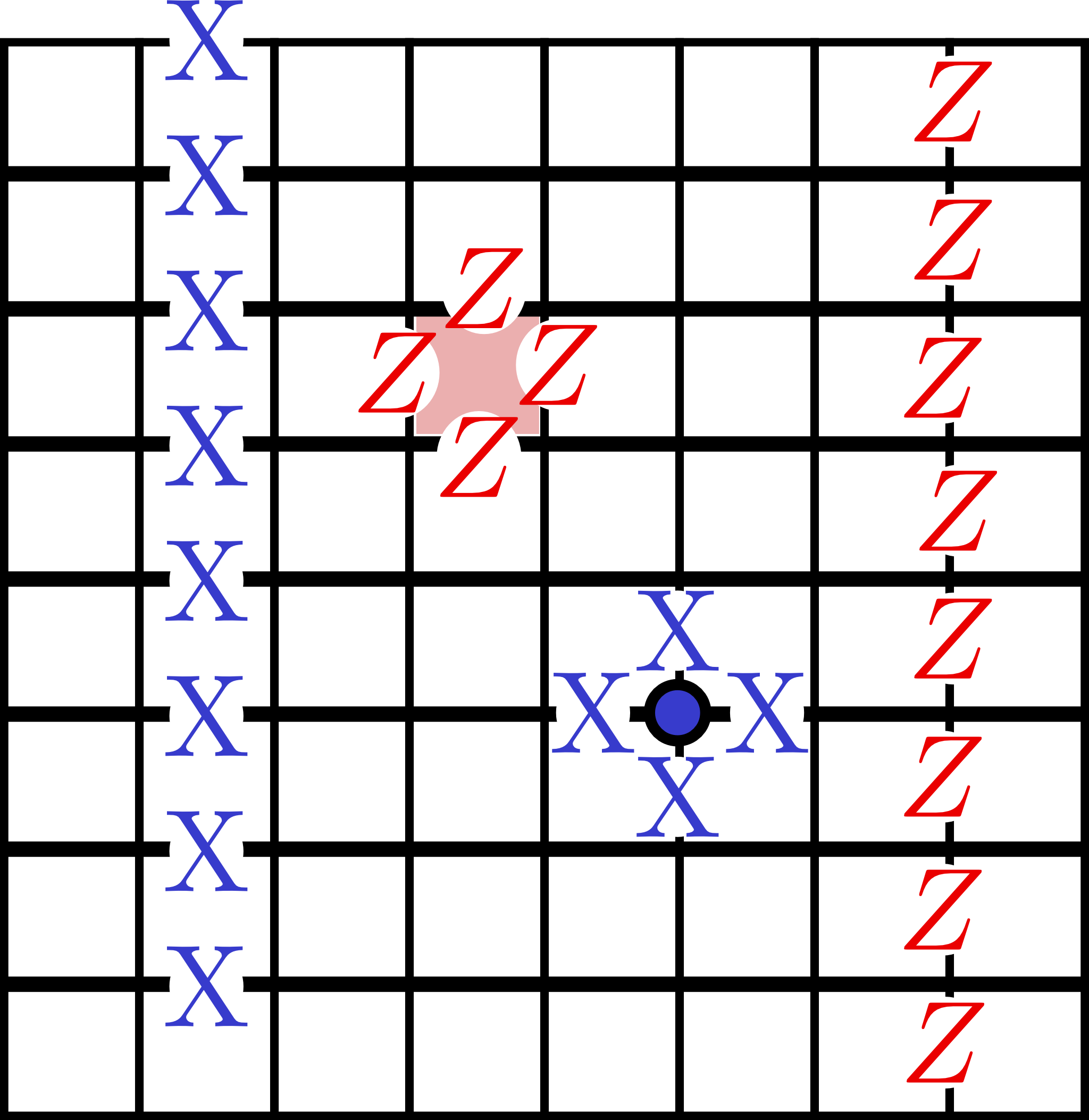}  
\end{center}
\caption{Each edge in the grid represents a physical qubit and opposite sides of the grid are identified by toric periodic boundary conditions.
Typical plaquette and vertex operators are depicted near the center.
Two vertical loop operators, $\bar{X}_1$ and $\bar{Z}_2$, which allow breaking the degeneracy are also presented.
One can take these to be the $X$ and $Z$ operators for the first and second logically encoded qubits respectively.
The complementary (anticommuting) operators are given by analogous horizontal loops.
}
\label{fig:ToricCode}
\end{figure}

In the toric code Hamiltonian, the physical qubits are located on the edges of a planar grid covering the 2D surface.
For concreteness and simplicity, we shall restrict to the case were the surface is a torus and the grid is an $N\times N$ square lattice (i.e.~$2N^2$ physical qubits).
The Hamiltonian is composed of commuting terms which are products of Pauli operators on different sites (it is a stabilizer Hamiltonian).
For each vertex $s$ of the grid, there is a star (or vertex) term $A_s = \prod_{j\in \text{star}(s)} X_j$ which is the product of $X$ operators over all the qubits of edges reaching $s$.
Analogously, for each face $p$ of the grid, there is a plaquette (or face) term $B_p = \prod_{j \in \text{boundary}(p)} Z_j$ which is the product of $Z$ operators over all the qubits of edges surrounding the face $p$.
Since each vertex and face have either 0 or 2 common edges, the terms $A_s$ and $B_p$ always commute.
Hence all terms of the toric code Hamiltonian 
\begin{equation}\label{eq:ToricHamiltonian}
 H=-\sum_s A_s -\sum_p B_p
\end{equation}
commute, and may be simultaneously diagonalized. 
Since $\prod_s A_s =I$ and $\prod_p B_p =I$, there are only $2 N^2 -2$ independent binary quantum numbers asociated to these terms (stabilizer operators) and each valid configuration determines a subspace of dimension $4$.
Due to this, violations of plaquette (vertex) conditions $A_s\ket{\psi} = \ket{\psi}$  ($B_p\ket{\psi} = \ket{\psi}$) always come in respective pairs. Following usual nomenclature, virtual particles called vertex (plaquette) anyons are respectively associated to these excitations.
The set of stabilizers may be completed with a pair of logical observables consisting of the product of $Z$ ($X$) operators along non contractible loops on the lattice (dual lattice), which may not be expressed as a product of plaquette (star) terms as illustrated in figure \ref{fig:ToricCode}.
Together with the set of Hamiltonian stabilizers any commuting pair of these four logical operators ($\bar{X}_1, \bar{X}_2, \bar{Z}_1$ and $\bar{Z}_2$) uniquely determine the state.

A stated prerequisite for using the toric code as a protecting Hamiltonian is that the energy splitting of the ground space due to Hamiltonian perturbations should be small.
This is argued through the use of degenerate perturbation theory and the fact that it only gives non-zero splitting when the order taken is at least the lattice width/height, claiming an exponential suppression of perturbations in the ground space.

The interaction terms in this Hamiltonian may be used as the syndrome measurements of an error correcting code with the desirable property that they are all geometrically local.
Such codes provide a way of obtaining a fault tolerance threshold without requiring the use of concatenated quantum error correction.
In this case, increasing the lattice size allows periodic measurements to suppress the effect of errors up to any desired accuracy \cite{dennis-2002} provided the accumulated error probability between measurements is below a certain threshold.

We will briefly review how error syndromes are interpreted and corrected, making the simplifying assumption that the error syndromes are measured perfectly. 
These syndromes, i.e.~measurements of the stabilizers, reveal the presence of any anyons on the lattice, but do not distinguish between them, so it is up to us to determine how these anyons should be paired up in order to annihilate them. For each of the two kinds of anyon, the error correcting procedure will pair up the anyons and annihilate them by applying a connecting string of operators on them.
If the connections performed and the actual origin of the anyons form topologically trivial loops (contractible loops), the error correction will have been successful.
If however, the actual error pathways, together with the connections performed by the error correction procedure complete one, or an odd number of, non-trivial loops, then a logical error will have been implemented.

\begin{figure}
\begin{center}
	\includegraphics[width=0.22\textwidth]{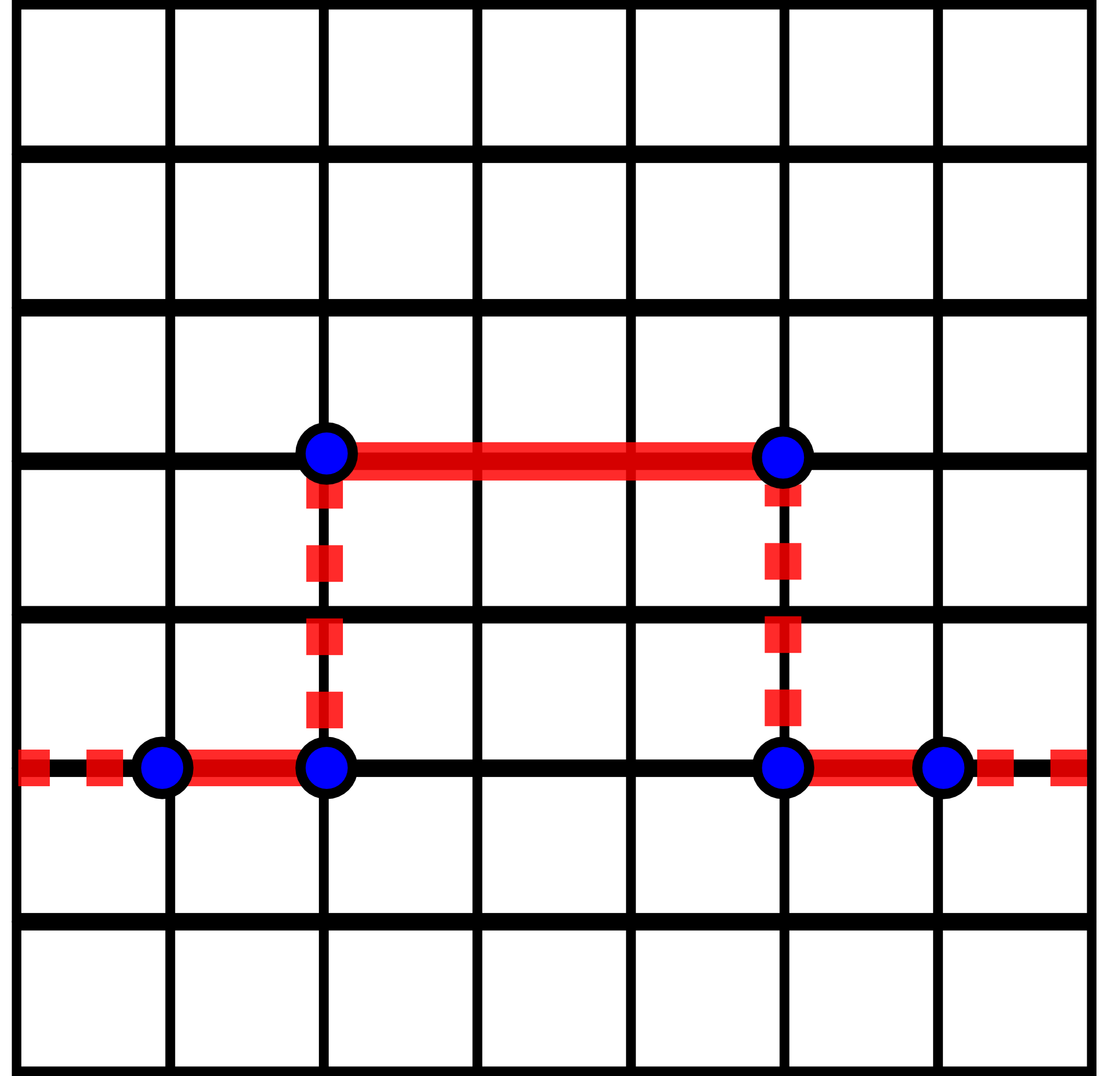}  
\end{center}
\caption{
Illustration of a possible configuration of three vertex anyon pairs (small circles).
Segments indicate possible qubits where $Z$ rotations could be introduced in order to remove the anyons.
Solid and dotted segments illustrate the anyon matching arising from $l_1$-EC and  $l_{\infty}$-EC respectively.
Since together they complete a non-trivial loop, the matchings are logically inequivalent.
}
\label{fig:AnyonPairing}
\end{figure}

Different criteria for pairing anyons may lead to logically different results.
This is illustrated in figure \ref{fig:AnyonPairing}, where two different criteria are used to pair up six anyons.
In particular, if one of the criteria compensates the actual error path, allowing recovery of the initial state, the other will complete the error path into an undesired logical operation.

There are two correction protocols which we will consider, as they are expected to perform adequately when correcting a small proportion of randomly located errors.
The first, which we refer to as $l_1$-EC, consists of minimizing the sum of distances among paired anyons,
for which there is a polynomial time algorithm \cite{vaidya_1988}.
The second, $l_{\infty}$-EC minimizes the furthest distance among paired anyons.

\section{\label{ap:ProofOfTCDepolarization}
Full Depolarization of the Toric Code's Protected Subspace}
In the main body of the paper (Sec.~\ref{sec:ToricOrderN}), we gave a construction for a single logical error $X_1$, $X_2$, $Z_1$ or $Z_2$ to be applied with a probability exponentially close to $50\%$, independent of the model used for error correction. 
This is not sufficient to show that we get full depolarization of the two-qubit subspace because it is not automatically clear that all 4 logical errors can be introduced simultaneously in the same model; the problem being that crossing paths for anti-commuting operations do not necessarily have a well-defined phase, and the perfect state transfer operations can fail.
Indeed, if two non commuting anyon propagation paths of equal length cross at their midpoints, the amplitude corresponding to full propagation on both paths can be seen to be $0$ at times!
It is the aim of this section to extend the setting of (Sec.~\ref{sec:ToricOrderN}) to multiple logical errors while ensuring that the failure probability remain exponentially small with system size, thereby allowing a fully depolarizing map on the code-space with probability exponentially close to 1.

The basic idea behind this construction is that, for large systems, the propagation of the anyons is essentially ballistic. Hence, we can divide our lattice into sections, and ensure that the paths for anyons of different types only cross in regions where we can be (almost) guaranteed of the order in which the anyons pass through. It is then our task to bound the error probability.

Let us first consider the probability $p_s(t)$ of finding a propagated anyon at site $s$ after a propagation time $t$ which is given in \cite{kay_geometric_2005} as
\begin{equation}
p_s(t) = f(s;D,\sin^2 t) = \sin^{2 s}t \cos^{2 (D-s)}t\binom{D}{s},
\end{equation}
where $f$ is the binomial distribution function and $D$ is the propagation length (i.e. there are $D+1$ possible anyon sites in the path).
Here time has been normalized such that perfect transfer occurs at $t=\frac{\pi}{2}$.
Correspondingly, if $P$ is the perfect transfer Hamiltonian for vertex anyons and $\Pi_0$ is the projector onto the subspace with a unique anyon at the transfer start site, then
\begin{equation}\label{eq:TransferExpansion}
\begin{array}{c}
 e^{-i t P}\Pi_0 = \sum_{s} \alpha_s(t) Z^{\otimes s}\Pi_0 \\
  \text{ where }\abs{\alpha_s(t)}=\sqrt{p_s(t)}
\end{array}
\end{equation}
and $Z^{\otimes s}$ is the tensor product of $s$ consecutive $Z$ operators along the anyon propagation path.

We are now in condition to compare an the actual evolution imposed by two non commuting anyon propagations $\ket{\psi(t)}$ and an ordered idealization of it $\ket{\psi_{b;a}(t)}$
\begin{equation}
\begin{array}{rcl}
\ket{\psi(t)} &=& \U e^{-i t (P_a + P_b)} \U^\dagger\ket{\psi_0} \\
\ket{\psi_{b;a}(t)} &=& \U e^{-i t P_a}e^{-i t P_b} \U^\dagger\ket{\psi_0}.
\end{array}
\end{equation}
We will assume that the physical qubit corresponding to the crossing of both paths is between anyon sites $s_{a}-1$ and $s_{a}$ of the anyon path associated to $P_a$ and between anyon sites $s_{b}-1$ and $s_{b}$ of the anyon path associated to $P_b$.
Furthermore, we will assume $s_a \gg s_b$, where what is meant by $( \gg )$ will soon be made clear.
Under these conditions, we will see that $\ket{\psi(t)}$ and $\ket{\psi_{b;a}(t)}$ are almost equal (at least during the time period corresponding to perfect state transfer).

By definition, we have that $\branoopket{\psi(0)}{\psi_{b;a}(0)}=1$.
Let us now bound how fast this overlap can actually decay
\begin{equation}
 \frac{d ~ \branoopket{\psi(t)}{\psi_{b;a}(t)}}{d~t} = 
 i\bra{\psi(t)}[P_b, e^{-i t P_a}] e^{-i t P_b} \U^\dagger\ket{\psi_0}.
\end{equation}
This allows bounding
\begin{equation}\label{eq:DerivativeBound}
 \abs{\frac{d ~ \branoopket{\psi(t)}{\psi_{b;a}(t)}}{d~t}}  \leq 
  \norm{[P_b, e^{-i t P_a}] e^{-i t P_b} \U^\dagger\ket{\psi_0}} .
\end{equation}
Now let $\Pi^{(a)}_\varnothing$ and $\Pi^{(a)}_0$ be projectors onto the subspace with no anyons in the path of $P_a$ and the subspace where a single anyon is located at the initial site and define $\Pi^{(b)}_\varnothing$ and $\Pi^{(b)}_0$ analogously.
Recalling that $\ket{\psi_0}$ is a code state and our choice of rotation $\U$, we have
\begin{equation}\label{eq:ProjectorIdempotence}
\begin{array}{rcl}
 (\Pi^{(a)}_\varnothing + \Pi^{(a)}_0)\U^\dagger\ket{\psi_0} &=& \U^\dagger\ket{\psi_0}\\
 (\Pi^{(b)}_\varnothing + \Pi^{(b)}_0)\U^\dagger\ket{\psi_0} &=& \U^\dagger\ket{\psi_0}.
\end{array}
\end{equation}
Commuting these projectors and using the expansion (\ref{eq:TransferExpansion}) of the perfect transfer we may express the RHS of equation (\ref{eq:DerivativeBound}) by
\begin{equation}\label{eq:StateNormBound1}
 \norm{[P_b, \sum_{s} \alpha_s(t) Z^{\otimes s}\Pi^{(a)}_0] 
 \sum_{r} \alpha_{r}(t) X^{\otimes r}\Pi^{(b)}_0 \U^\dagger\ket{\psi_0}}
\end{equation}
There is only one possible non commuting term in $P_b$ and this only for $s \geq s_a$.
Furthermore, this term cancels for all but two terms in the sum over $s'$.
We may then rewrite (\ref{eq:StateNormBound1}) as
\begin{equation}\label{eq:StateNormBound2}
 \norm{ \begin{array}{l}
 2J_{s_b}~\sum_{s\geq s_a} \alpha_s(t) Z^{\otimes s}\Pi^{(a)}_0 \times\\
  \times\left(\alpha_{s_b}(t) X^{\otimes s_b-1} + \alpha_{s_b-1}(t) X^{\otimes s_b} \right) \Pi^{(b)}_0 \U^\dagger\ket{\psi_0}
 \end{array}}
 \end{equation}
Where $J_{s_b}$ is the strength of the term performing an anyon swap between sites $s_b$ and $s_b-1$.
Since each coefficient accompanies an orthogonal component of the state, we may recall the definition in (\ref{eq:TransferExpansion}) and rewrite (\ref{eq:StateNormBound2}) as
\begin{equation}\label{eq:StateNormBound3}
 2J_{s_b}\sqrt{\left[ p_{s_b-1}(t)+p_{s_b}(t)\right] \sum_{s\geq s_a} p_s(t)}\sin^2\epsilon,
\end{equation}
where $\sin^2\epsilon$ is the amplitude of $\Pi^{(a)}_0\Pi^{(b)}_0\U^\dagger\ket{\psi_0}$.
An exponentially small upper bound will now be given for the expression inside the square root .
\begin{equation}\label{eq:expBound1}
\begin{array}{cl}
& \left[ p_{s_b-1}(t)+p_{s_b}(t)\right] \sum_{s\geq s_a} p_s(t) \\
\leq &  \sum_{r\leq s_b} f(r,D,\sin^2 t) \sum_{s\geq s_a} f(s,D,\sin^2 t) \\
 = & F(s_b, D, \sin^2 t) F(D-s_a,D, \cos^2 t),
\end{array}
\end{equation}
where $F(k,N,p)=\sum_{i=0}^k f(i,N,p)$ is the cumulative binomial distribution function.
Assuming $\frac{s_b}{D} \leq \sin^2 t \leq \frac{s_a}{D}$ we may use Hoeffding's inequality \cite{hoeffding_probability_1963} to bound (\ref{eq:expBound1}) as
\begin{equation}\label{eq:expBound2}
 e^{-2\frac{(D\sin^2 t-s_b)^2}{D}} e^{-2\frac{(D\sin^2 t- s_a)^2}{D}} \leq e^{-\frac{(s_a-s_b)^2}{D}},
\end{equation}
with equality holding for $\sin^2 t = \frac{s_a + s_b}{2D}$.
In turn, a tighter bound can be obtained by using Hoeffding's inequality on a single factor of (\ref{eq:expBound1}) when $\sin^2 t \geq \frac{s_a}{D}$  or $\frac{s_b}{D} \geq \sin^2 t$.

Taking $D = N/2-1$ as in Sec.~\ref{sec:ToricOrderN} and $s-r \geq s_a-s_b \geq D/6$ for instance, the obtained upper bound becomes exponentially small in $N$.
In turn, the derivative (\ref{eq:DerivativeBound}) is exponentially small, meaning that the actual evolution is approximated by the ordered evolution with exponentially good precision in $N$.

\begin{figure}[!tbh]
\begin{center}
\includegraphics[width=0.35\textwidth]{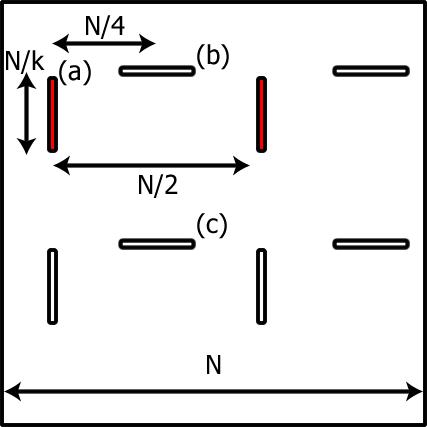}
\end{center}
\vspace{-0.5cm}
\caption{In an $N\times N$ lattice, there are two sets of $N/k$ rows ($k\sim O(1)$) and two sets of columns and rows, each of which corresponds to the construction of (Sec.~\ref{sec:ToricOrderN}) for a different error type ($\bar{X}_1$, $\bar{Z}_2$ are introduced by columns starting at horizontal stripes and $\bar{Z}_1$  and $\bar{X}_2$ are introduced by rows starting from vertical stripes).}\label{fig:toric_setup}
\vspace{-0.5cm}
\end{figure}

We have formally proven that for two non commuting anyon propagation paths which intersect with a sufficiently large offset (i.e.~$\geq D/6$) the evolution can be accurately approximated by ordered anyon propagation.
There is no obstacle in generalizing this result to many such anyon paths, as required to introduce logical errors with high probability.
Some leading factors of order $N^2$ appear but crucial factors remain exponentially decreasing in $N$.

In Fig.~\ref{fig:toric_setup}, we illustrate a configuration allowing the simultaneous introduction of all possible logical errors by anyon propagation within the lattice.
The marked stripes of width $N/k$ indicate locations where perpendicular anyon propagations begin or end.
The perturbation to be introduced is chosen randomly as in Sec.~\ref{sec:ToricOrderN}, such that each propagation row/column starts with equal probability in either of each pair of opposing stripes. 
Taking $k$ fixed allows sufficiently many repetitions of the single row/column construction that the probability of introducing each type of logical error approaches $\half$ exponentially fast with $N$. 
Thus, after a perturbed evolution for time $t_f = O(N)$, and a final application of an arbitrary error correcting protocol based on unperturbed syndrome measurement, the resulting state is exponentially close to the maximally mixed state $\identity/4$ of the code-space.

\bibliography{HamiltonianPerturbed}

\end{document}